\documentclass[aps,prb,preprint,a4paper,showpacs]{revtex4-1}
\usepackage{graphicx}
\usepackage{amsmath}
\usepackage{hyperref}
\begin{document}
\title{Low non-linearity spin-torque oscillations driven by ferromagnetic nanocontacts} 
\author{Muftah Al-Mahdawi}
\email{mahdawi@ecei.tohoku.ac.jp}
\author{Yusuke Toda}
\author{Yohei Shiokawa}
\author{Masashi Sahashi}
\affiliation{Department of Electronic Engineering, Tohoku University, Sendai 980-8579, Japan}
\date{\today}
\begin{abstract}
Spin-torque oscillators are strong candidates as nano-scale microwave generators and detectors. However, because of large amplitude-phase coupling (non-linearity), phase noise is enhanced over other linear auto-oscillators. One way to reduce nonlinearity is to use ferromagnetic layers as a resonator and excite them at localized spots, making a resonator-excitor pair. We investigated the excitation of oscillations in dipole-coupled ferromagnetic layers, driven by localized current at ferromagnetic nano-contacts. Oscillations possessed properties of optical-mode spin-waves and at low field ($\approx$200 Oe) had high frequency (15 GHz), a moderate precession amplitude (2--3$^\circ$), and a narrow spectral linewidth ($<3$ MHz) due to localized excitation at nano-contacts. Micromagnetic simulation showed emission of resonator's characteristic optical-mode spin-waves from disturbances generated by domain-wall oscillations at nano-contacts.
\end{abstract}
\maketitle
\section{Introduction}
Transfer of angular momentum from a dc spin-polarized current to a nano-scale ferromagnet (FM) exerts an anti-damping torque, Spin-Transfer Torque (STT), that can compensate intrinsic damping torque and induce stable precession of magnetization.\cite{slonczewski1997} When combined with a magnetoresistance effect, like giant magnetoresistance (GMR) or tunneling magnetoresistance (TMR), high-frequency voltage oscillations are emitted, producing Spin-Torque Oscillations (STO).\cite{kiselev2003} The same structures can also rectify injected ac voltage at resonance,\cite{tulapurkar2005,miwa2014} and off-resonance\cite{fu2012}. Such microwave nano-oscillators/detectors are sought after for applications like inter-/intra-chip communication, imaging\cite{fu2014} and non-destructive testing, and lab-on-chip sensors. STOs are suitable candidates for such a role. However, low output power, the trade-off between power and frequency, and weak coherence of oscillations hinder their applications compared with their semiconductor counterparts. The large nonlinear phase-amplitude coupling\cite{kim2008} produces a dilemma between large precession amplitude and small linewidth. Also, frequency locking of STOs to external reference signal becomes hindered.\cite{georges2008-a} Such a dilemma can be overcome by the separation of STOs into an excitation source and a resonating element, so that precession frequency and amplitude will be set by the resonator design, not by the driving STT. Then the linewidth will be reduced considerably.\\
It was shown that non-uniform current density in TMR-STO resulted in an increase the amplitude of generated precession,\cite{houssameddine2008,maehara2013,maehara2013-a,maehara2014} and reduction in linewidth.\cite{houssameddine2008,houssameddine2009,devolder2009,kudo2009} However, the origin is still not clear, and the fabrication process is not well understood or easily reproducible. On the other hand, the Ion-Assisted Oxidation (IAO) of ultra-thin aluminum reproducibly was used to fabricate a 1-nm-thin alumina Nano-Oxide Layer (NOL) with direct 2-nm Nano-Contacts (NCs) between FM layers.\cite{fukuzawa2004,fuke2007,takagishi2009,shiokawa2011,yuasa2013} In NOL-based STOs, moderate power and narrow linewidths were reported,\cite{endo2009,suzuki2009,suzuki2011,doi2011} with oscillation behavior similar to low-TMR-STO.\cite{al-mahdawi2011} There is evidence for the presence of FM metallic NCs by magnetoresistance and transport properties,\cite{fuke2007} transmission-electron micrographs,\cite{takagishi2009,yuasa2013,al-mahdawi2014} and conductive atomic force microscopy.\cite{takagishi2009,shiokawa2011,al-mahdawi2014} However, the measured Nano-Contacts Magnetoresistance (NCMR) ratios are far below expectation compared with scattering from confined Domain-Walls (DW),\cite{tagirov2001,sato2009,imamura2011} mostly due to the presence of oxygen and non-magnetic impurities.\cite{shiokawa2011,shiokawa2015,al-mahdawi2014}\\
In this paper, we propose that the localized precession of DWs at NCs work as excitors of spin-waves in FM layers. This makes frequency completely determined by resonator's designed eigen-frequency, regardless of the mechanism of excitors. The loss of frequency tunability reduces non-linearity and linewidth considerably. After the experimental and simulation description, we present the excited modes in the chosen resonator, then we discuss the reduced non-linearity of NCMR-STO. For the resonator, we used a nano-pillar with two free FM layers, where magnetostatic dipolar field provides inter-layer coupling with two coupled-oscillations characteristic modes.\cite{grunberg1980} The dynamics of coupled free FM layers are of practical interest for high-frequency emission at low applied field ($<$ 500 Oe), linewidth narrowing, and doubling of magnetization precession frequency in resistance oscillations.\cite{seki2010,moriyama2012,braganca2013,nagasawa2014} Most importantly, the loss of current-tunability of frequency can be compensated by the controllability of inter-layer relative angle.\\
\section{Experiment and simulation}
The film stack with designed thickness in nm was: thermally-oxidized silicon substrate/electrode layer(Ta 5/Cu 200/Ta 40/chemical-mechanical polishing)/milling 5/Ta 3/Ru 2/Fe$_{50}$Co$_{50}$ 5/Al 1.3/IAO 20 seconds exposure time/Al 0.3/Fe$_{50}$Co$_{50}$ 5/capping (Cu 10/Ru 10). The film was deposited by magnetron and ion-beam sputtering in the chambers described before.\cite{shiokawa2011} Subsequently, films were vacuum-annealed at 270$^\circ$C and 400$^\circ$C for 1.5 hours each with 10-kOe magnetic field. The choice of capping material and additional metal Al insertion were optimized with annealing process for lower Resistance-Area (RA) product and enhanced MR ratio, based on previous work.\cite{shiokawa2011,shiokawa2015} Current-Perpendicular-to-Plane (CPP) pillars of elliptical cross-section were micro-fabricated by Ar$^+$ ion-milling and electron-beam lithography. RA was found from the slope of 4-probe dc resistance vs.~area inverse (R-1/A) line, and compared with Current-In-Plane-Tunneling (CIPT) measurement of unpatterned films.\\
The STO microwave emissions were measured by a 26-GHz spectrum analyzer under biasing from a bias-T, and scattering at the measurement probe was measured by a network analyzer. Geometry, angles and coordinates definitions are summarized in Fig.~\ref{fig:RH}(a). The positive current was defined to be electrons flowing up. We are presenting the detailed measurements of a 320 nm $\times$ 160 nm pillar at $\xi=60^\circ$, although results presented later were qualitatively similar among samples.\\
Micromagnetic simulation was done by Nmag, a finite-element-method simulator.\cite{fischbacher2007} Calculation geometry consisted of two ellipses same as experimental design separated by 1 nm, and NCs were included as 1-nm-radius cylindrical contacts between the two layers. Inter-layer dipolar field was included through demagnetization field calculation. Material parameters are: stiffness constant of $2.3\times10^{-10}$ erg/cm$^3$, saturation magnetization of 1930 emu/cm$^3$,\cite{miyake2013} with Gilbert damping constant of 0.02, and an unphysical spin polarization of 100\%. Mesh size away from NCs was set to 5 nm, and it changed to 0.7 nm inside NCs. For hysteresis loops, $\xi$-dependence was calculated for 20 randomly-placed NCs. For qualitative understanding of STO dynamics, we compared the cases of zero and four NCs, under the experimental conditions of 250-Oe field applied at $\xi=60^\circ$. The current profile was approximated to be confined in NCs with confinement extending 1 nm away from middle of NC into FM layers, as most of the voltage drop will be on this region,\cite{naidyuk2005} although more accurate representation is needed.\cite{strelkov2011} The total current was +17.5 mA and the current distribution was calculated by assuming that a single NC and tunnel barrier resistances are 600 $\Omega$ and 500 $\Omega$, respectively.\cite{takagishi2009}\\
\section{Results and discussion}
\begin{figure}
	\caption{(a) A schematic of the elliptical pillar geometry and definitions of coordinates and angles. (b) Two-probe resistance vs.~field applied at $\xi=0^\circ$ and $60^\circ$. Inset shows the bias dependence of differential resistance in parallel magnetizations state. (c) Normalized MR found from micromagnetics simulation, for the cases of no NCs (dashed lines), and with 20 NCs (solid lines).}
	\label{fig:RH}
	\includegraphics[width=0.7\textwidth]{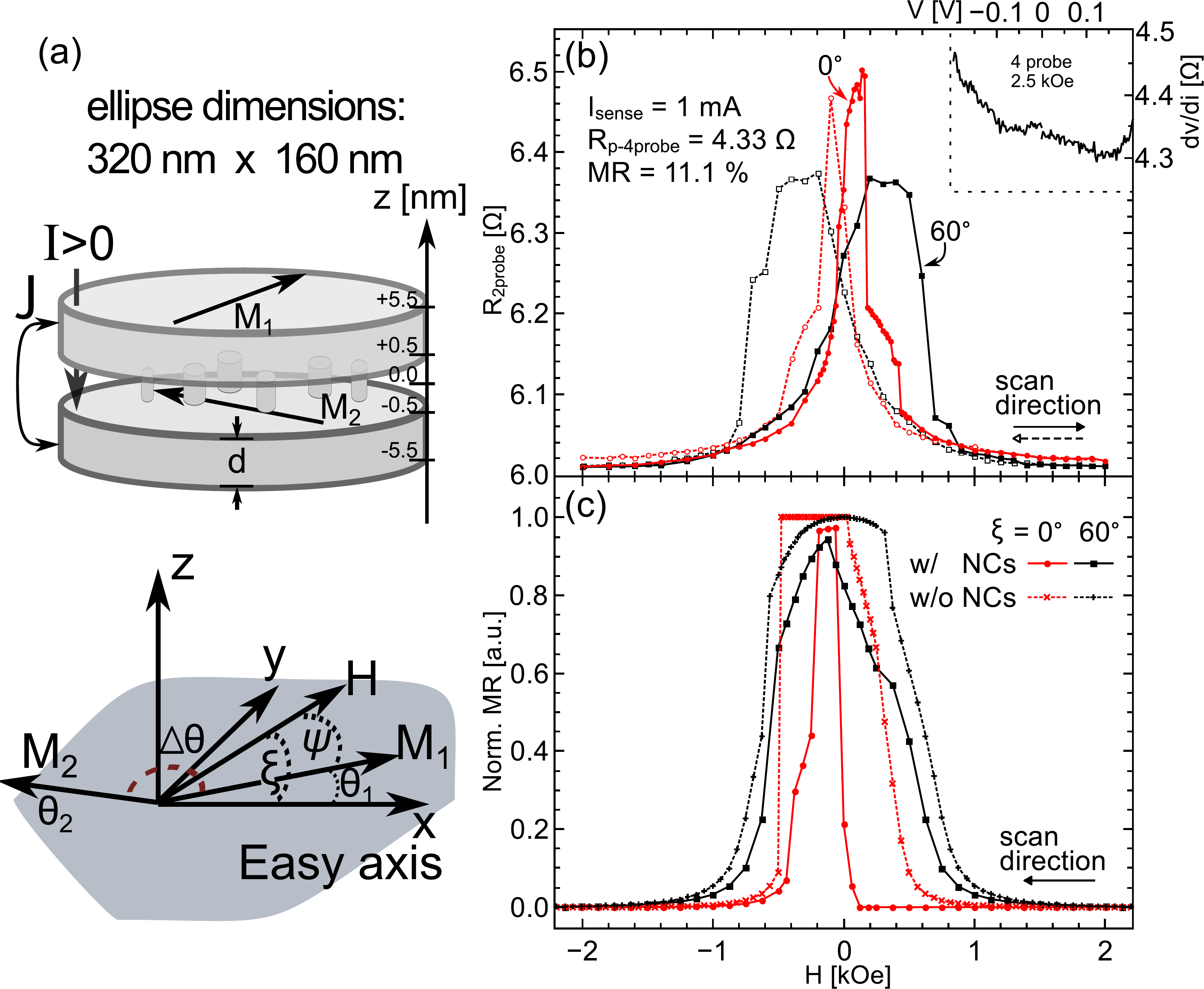}
\end{figure}

\begin{figure}
	\caption{(a) Representative oscillation power spectrum  at $\xi=60^\circ$ with Lorentzian peak fitting (solid line). (b) The acoustic and optical modes of coupled oscillations were observed near 3 GHz and 15 GHz, respectively. (c) Current dependencies of oscillation characteristics. Same symbols in $f_{\mathrm{osc}}$ and $\Delta f$ panels correspond to each other.}
	\label{fig:osc}
	\includegraphics[width=0.6\textwidth]{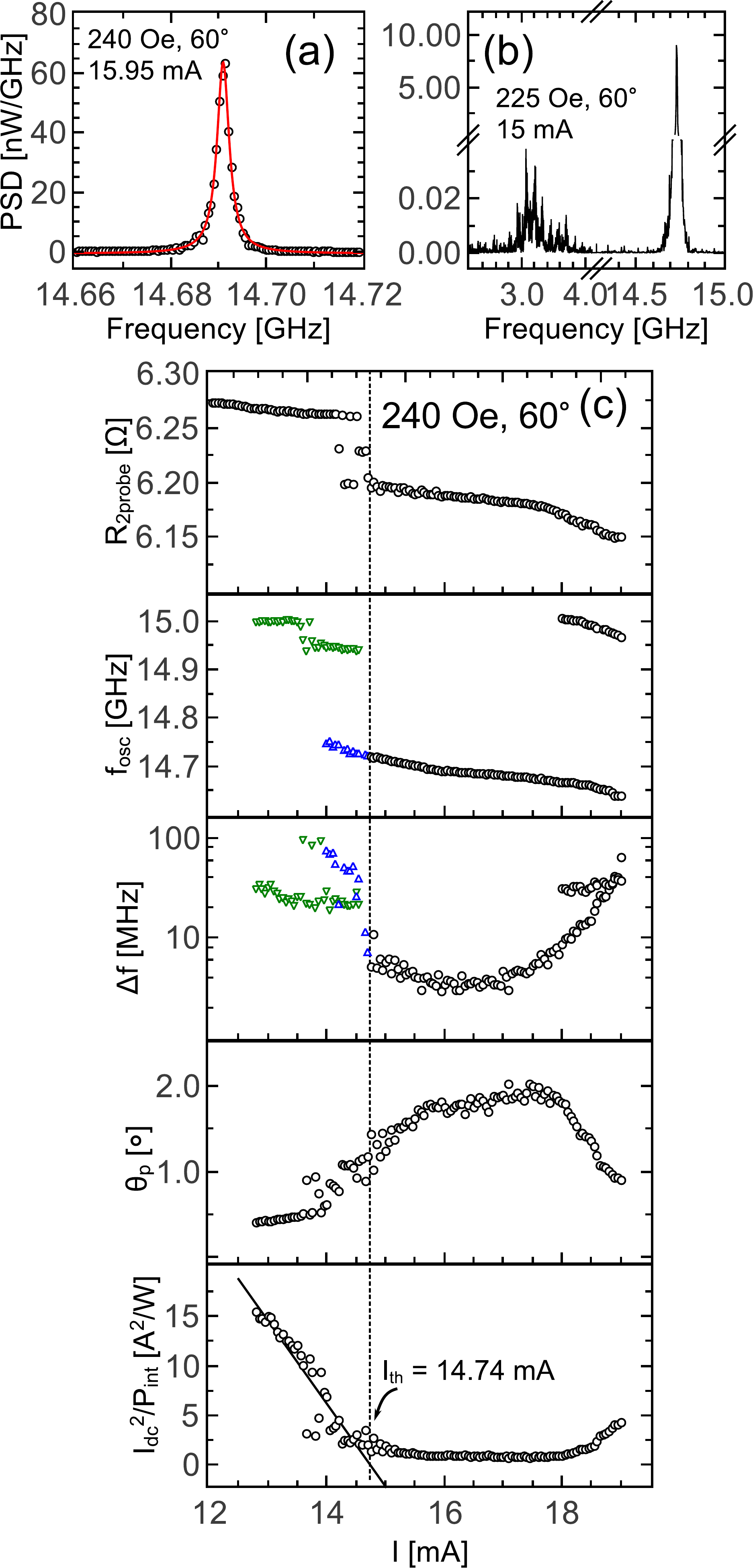}
\end{figure}

\begin{figure}
	\caption{(a--c) Comparison of micromagnetic dynamics with (top part) or without (bottom part) NCs at $H = 250$ Oe, $\xi=60^\circ$, $\mathrm{I_{dc}}=17.5$ mA . (a) Normalized magnetization's y-component of the total system ($m_y$). (b) Magnetization angle of top and bottom layers at $(x,y)=(0,0)$. (c) Spectrum of $m_y$ transformed from 250-ns (15-ns) duration of dynamics without (with) NCs. (d) Time snapshots with the color representing $\left[m_y(\mathbf{r},t)-m_y(\mathbf{r},0)\right]$. Optical-mode spin-waves are emitted from localized excitation at NCs. (e) The origin of localized excitation is a confined domain-wall that is oscillating at 250-GHz. (f) Power profile of localized precession around the top-right NC.}
	\label{fig:oscsim}
	\includegraphics[width=0.8\textwidth]{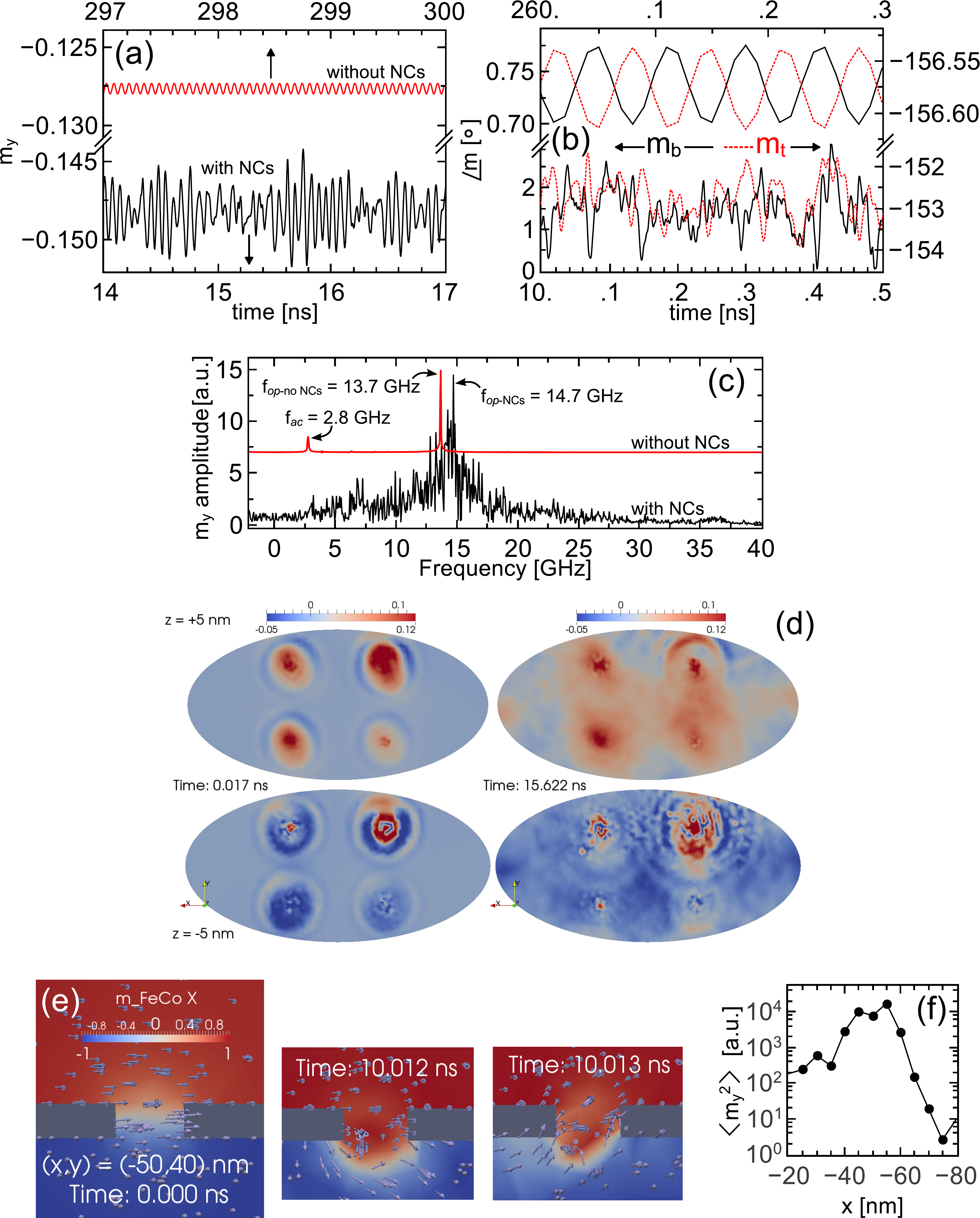}
\end{figure}

RA found from the slope of 4-probe R-1/A line, and CIPT measurements were 0.2 and 0.3 $\Omega\cdot\mu\mathrm{m^2}$, respectively. The bias dependence of 4-probe differential resistance at parallel magnetization state (inset in Fig.~\ref{fig:RH}(b)) was relatively flat. The resistance temperature dependence of similarly conditioned films also showed metallic-transport character. This indicates that conduction is dominated by transport through NCs and not by tunneling through oxide barrier. Previously, high-temperature annealing ($>380^\circ$C) was hindered by manganese diffusion from pinning antiferromagnet towards NOL.\cite{shiokawa2011} Better NOL barrier quality and purer NCs were obtained in this report by using manganese-free structure for higher annealing temperature, in addition to insertion and capping layers optimizations.\cite{shiokawa2015}\\
Figure \ref{fig:RH}(b) shows the two-probe resistance vs.~magnetic field (R-H) applied at $\xi=0^\circ$ and $60^\circ$ measured at the same position as STO measurements. From the switching fields of easy-axis R-H, interlayer dipolar coupling field ($H_{ic}$) is estimated at 400 Oe. It is in agreement with the estimation of cross-demagnetization,\cite{dmytriiev2010} $H_{ic}=4\pi\rho_{12}M_s = 433$ Oe, where $M_s$=1750 emu/cm$^3$ is the measured saturation magnetization, and $\rho_{12}=0.0197$ is the cross-demagnetization factor. This dipole inter-layer coupling can be considered equivalent to the usual bilinear coupling through a metal spacer ($J=-2d M_s H_{ic}=-0.7$ erg/cm$^2$,\cite{grimsditch1996} defined negative for anti-parallel coupling). The contribution from coupling through NCs and spacer roughness to magnetostatic energy can be neglected. Ferromagnetic coupling was found to be small from free-layer magnetization-loop shift of unpatterned spin-valve films (IrMn/FeCo(Pinned)/NOL-NCs/FeCo(Free)), with $J_{\mathrm{NCs}} = 0.01-0.02$ erg/cm$^2$ .\cite{shiokawa2011}\\
Micromagnetic simulation reproduced static R-H, $H_{ic}$ estimation, and the reduction of AP-to-P plateau width by NCs (Fig.~\ref{fig:RH}(c)). We chose for oscillation measurements the pillar that had the closest R-H curve to micromagnetic simulation, which had 11.1\% MR ratio and 0.17 $\Omega\mu\mathrm{m^2}$ RA product. Due to large pillar size, uniform rotational switching was not reproducible for field applied along easy axis ($\xi=0^\circ$). At tilted angles, the magnetization rotated as a single domain. Thus we are presenting tilted angles results of oscillations (Fig.~\ref{fig:osc}).\\
Largest power microwave oscillations were observed for $\xi=60^\circ$ at $\approx$15 GHz when applying high currents. Sample power spectrum with a Lorentzian peak fitting is shown in Fig.~\ref{fig:osc}(a). There is a drop in resistance at $\mathrm{I_{dc}}$ = 14.7 mA accompanied with a jump in oscillation frequency, $f_{\mathrm{osc}}$, a narrowing in full-width-at-half-maximum, $\Delta f$, and increase in integrated power, $\mathrm{P_{int}}$, indicating a change into auto-oscillation mode,\cite{slavin2009} with a mechanism similar to STOs based on pin-hole tunnel junctions (Fig.~\ref{fig:osc}(c)).\cite{houssameddine2008} Linear fits to normalized inverse power, $1/p$, at sub-threshold gave a threshold current, $\mathrm{I_{th}}$ of 14.74 mA. The presense of two frequency branches at sub-threshold and high-current regions can be ascribed to edge and center modes in elliptical geometries.\cite{deac2008} The highest oscillation power is 0.4 nW (1.6 nW if corrected for impedance mismatch) giving a precession amplitude ($\theta_p$) of 2--3$^\circ$, whereas the lowest $\Delta f$ is 3 MHz corresponding to a quality factor of 5000.\\
Regarding the excited oscillations, the possible coupled oscillations or spin-waves in two layers of free spins are the optical (anti-phase) mode (OM) and the acoustic (in-phase) mode (AM).\cite{grunberg1980} Modes frequencies can be found from the solution to coupled Bloch equations of the two layers with effective field determined from the free energy.\cite{cochran1990,grimsditch1996,zivieri2000} We considered only the main contributions of Zeeman energy, film demagnetization, and interlayer dipolar coupling. The optical and acoustic eigen-frequencies of an in-plane magnetization precession can be simplified to:
\begin{widetext}
\begin{subequations}\label{eq:freq}
\begin{align}
\left( \frac{f_{\mathrm{ac}}}{\gamma/2\pi} \right)^2 =&\left( H \cos\psi + 4\pi M_s - 2 H_{ic} \left(\cos\Delta\theta + 1 \right)\right) \left( H\cos\psi \right) + 8 H_{ic}^2\cos\Delta\theta, \label{subeq:f_ac}\\
\left( \frac{f_{\mathrm{op}}}{\gamma/2\pi} \right)^2 =&\left( H \cos\psi + 4\pi M_s - 2 H_{ic} \left(\cos\Delta\theta + 1 \right)\right) \left( H\cos\psi - 4 H_{ic}\cos\Delta\theta \right), \label{subeq:f_op}
\end{align}
\end{subequations}
\end{widetext}
where $\gamma/2\pi = 2.8$ MHz/Oe is the gyromagnetic ratio, and other symbols are defined in Fig.~\ref{fig:RH}(a). We confirmed the presence of the weaker-amplitude AM (Fig.~\ref{fig:osc}(b)). Using $H_{ic}$ = 400 Oe and $\Delta\theta=130$--$150^\circ$ from measured R-H and micromagnetic simulation results in $f_{\mathrm{op}}$ of 14.0--16.1 GHz and $f_{\mathrm{ac}}$ of 3.9--3.5 GHz, which agrees with the observed spectrum. The frequency of OM depends mostly on the coupling strength and relative angle between the layers (the last term on right in Eq.~\ref{subeq:f_op}). The weak dependence of $f_{\mathrm{osc}}$ against $\mathrm{I_{dc}}$ and H ($\approx -1.3$ MHz/Oe not shown) supports that $f_{\mathrm{osc}}$ is determined mainly by excitation of an OM spin-wave. The maximum $f_{\mathrm{osc}}(H=0,\Delta\theta=180^\circ)$ from other devices was 17.8 GHz which corresponds to $H_{ic}$ = 460 Oe, in agreement with the estimation from the corresponding R-H curve. To increase the oscillation frequency, we reduced the size of elliptical pillars to 160 nm $\times$ 80 nm. At $\mathrm{I_{dc}}$ = 2.8 mA, H = 185 Oe, and $\xi=50^\circ$, the resulting oscillations had $f_{\mathrm{osc}}$, $df/dI_\mathrm{dc}$, and $\Delta f$ of 23.3 GHz, $<4$ MHz/mA, and 1.3-MHz linewidth, respectively. The corresponding quality factor is more than 17,000.\\
Although the presented frequency of OM is higher than AM, the measured and simulated peak intensities of OM are much larger (Figs.~\ref{fig:osc}(b) and \ref{fig:oscsim}(a,c)). This is due to two reasons. In OM, anti-phase dynamics maximize dynamic MR change.\cite{moriyama2012,braganca2013} In comparison, OM amplitude as measured by Brillouin light scattering is reduced due to canceling contributions to light scattering cross-section.\cite{grimsditch1996} Secondly, the energy required to excite OM is smaller in anti-coupled ($J<0$) harmonic oscillators. The average energy difference between AM and OM with equal precession amplitudes is: $\left\langle\mathrm{E_{ac}}\right\rangle-\left\langle\mathrm{E_{op}}\right\rangle = -J\left( \delta\mathbf{m}_1\cdot\delta\mathbf{m}_2 \right)$, where $\delta\mathbf{m}_1\cdot\delta\mathbf{m}_2$ is the characteristic-mode's dimensionless power. For $J<0$, excitation of AM requires higher energy than same-amplitude OM.\\
It should be noted that Eqs.~\ref{eq:freq} were derived for infinite-wavelength limit (\emph{i.e.}~wave-vector-thickness product $qd\approx 0$). Quantitative corrections due to dynamic dipolar coupling between propagating spin-waves cannot be ignored because $qd = 1.74$ from simulation presented later.\cite{cochran1990,grimsditch1996,zivieri2000} But due to experimental uncertainty in determining $H_{ic} \cos\Delta\theta$, exact quantitative comparison becomes difficult, and the main conclusions are not changed.\\
The linewidth broadening of STOs compared to linear auto-oscillators is understood to be due to amplitude-phase coupling,\cite{kim2008} which is expressed by the nonlinearity parameter ($\nu$).\cite{tiberkevich2008} The nonlinearity of presented results is $\nu = (I_\mathrm{dc}/\Gamma_g)(df/dI_\mathrm{dc}) \approx-0.16$. The natural FMR linewidth ($\Gamma_g = 934$ MHz) is obtained from linear extrapolation of $\Delta f$ to zero current at sub-threshold,\cite{georges2009} and the agility of oscillation frequency in current ($df/dI_\mathrm{dc}$) was -9.6 MHz/mA. This non-linearity is one order of magnitude smaller than other reported values.\cite{georges2009,kudo2009,bianchini2010} Because the nonlinearity is very small, the sudden change of oscillation into a single mode at threshold hinders the applicability of determining $\nu$ from $\Delta f$-$1/p$ plot.\cite{kudo2009} The $\Delta f$-$1/p$ slopes were 3.8 and 19.4 $\mathrm{MHz}/(\mathrm{mA}^{2}\cdot\mu\mathrm{W})$ for above-threshold and below-threshold regimes, respectively.\\
NCMR-STO usually showed relatively small agility (16--18 MHz/mA)\cite{endo2009,suzuki2011,al-mahdawi2011} compared with other TMR-STOs, leading to smaller nonlinearity and narrow linewidth. Possible reasons for lowered agility and nonlinearity in this report can be the optimized fabrication process with purer NCs,\cite{shiokawa2011,shiokawa2015} the coupled oscillations of two layers,\cite{gusakova2011,moriyama2012,nagasawa2014} and the tilted magnetization angle away from easy axis.\cite{mizushima2009} However, the loss of agility and small linewidths were obtained for various angles and frequencies, which indicates that the improved purity of NCs is the main factor.\\
We used micromagnetic simulation to show how nano-magnets with NCs worked as a resonator-excitor pair. Results shown in Fig.~\ref{fig:oscsim} agree reasonably with experimental data. In the case of no NCs (upper part in Figs.~\ref{fig:oscsim}(a--c)) oscillation frequency is similar to the calculated and measured ones, with the optical mode being the dominant component. But precession amplitude ($\theta_p=0.05^\circ$) is very small compared with the experimental value (2--3$^\circ$). With the insertion of four NCs, optical spin-waves were emitted from NCs (Fig.~\ref{fig:oscsim}(d)), and increased $\theta_p$ to 2$^\circ$ (lower part in Fig.~\ref{fig:oscsim}(a--c)).
The origin of perturbation near NCs is of similar origin to previous reports.\cite{arai2012,matsushita2010-a} The domain-wall is pushed outside NC-region into the ferromagnetic layer and starts to oscillate at high frequency between N\'eel and Bloch walls (250 GHz for the chosen geometry and current density)(Fig.~\ref{fig:oscsim}(e)). These very high frequency oscillations were localized up to 5--10 nm away from NC (Fig.~\ref{fig:oscsim}(f)). The localized precession acts as a point source that generates spin-waves propagating radially at the characteristic mode of the system, which is an optical spin-wave.\\
The implication on the nature of current-induced dynamics is that magnetization precession is not induced by STT directly. In Fig.~\ref{fig:osc}(c), at first going above I$_\mathrm{th}$, localized STT increases amplitude, compensates damping around NCs and increases local precession amplitude. When local precession amplitude saturates, at I = 15.6 mA, increase of current and STT will not change $\theta_p$, leading to loss of agility and narrowing of linewidth. At this stable regime, magnetic layers act as a resonator that is excited by the energy coming from the point sources at NCs. This makes the presented oscillator similar to a classical auto-oscillator, and results in a considerable reduction in linewidth.\\
\section{Conclusion}
In conclusion, we presented the measured spin-torque-driven oscillations of a spin-torque oscillator with nano-contacts between two free ferromagnetic layers coupled antiferrmagnetically with dipolar-field. Resulting oscillation character agrees with propagating optical-mode spin-waves. In micromagnetic simulation, inclusion of NCs with localized current density showed that NCs work as point sources, and optical-mode spin-waves were excited in the ferromagnetic layers. The ferromagnetic layers act as a resonator that is decoupled from mechanism of excitation point sources. Since oscillation is not driven by spin-transfer torque, linewidth decreased.\\
Although we studied dual ferromagnetic layers in this report, same effect should be the origin for the small non-linearity and linewidth narrowing common in NCMR and pin-hole-TMR STOs.\cite{endo2009,suzuki2011,al-mahdawi2011,houssameddine2008,houssameddine2009} So, by utilizing NCs and using a magnetic resonator that has a large oscillation amplitude, \emph{e.g.}~tilted-anisotropy ferromagnets,\cite{nguyen2012} we expect to increase both power and quality factor for applications.
\begin{acknowledgments}
Authors would like to thank Dr.~Andrei Slavin for his insightful comments and discussion. This work was supported by SCOPE program (contract No.~0159-0058) from Ministry of Internal Affairs and Communications.
\end{acknowledgments}
\bibliographystyle{apsrev4-1}
\bibliography{refs}

\begin{thebibliography}{53}%
\makeatletter
\providecommand \@ifxundefined [1]{%
 \@ifx{#1\undefined}
}%
\providecommand \@ifnum [1]{%
 \ifnum #1\expandafter \@firstoftwo
 \else \expandafter \@secondoftwo
 \fi
}%
\providecommand \@ifx [1]{%
 \ifx #1\expandafter \@firstoftwo
 \else \expandafter \@secondoftwo
 \fi
}%
\providecommand \natexlab [1]{#1}%
\providecommand \enquote  [1]{``#1''}%
\providecommand \bibnamefont  [1]{#1}%
\providecommand \bibfnamefont [1]{#1}%
\providecommand \citenamefont [1]{#1}%
\providecommand \href@noop [0]{\@secondoftwo}%
\providecommand \href [0]{\begingroup \@sanitize@url \@href}%
\providecommand \@href[1]{\@@startlink{#1}\@@href}%
\providecommand \@@href[1]{\endgroup#1\@@endlink}%
\providecommand \@sanitize@url [0]{\catcode `\\12\catcode `\$12\catcode
  `\&12\catcode `\#12\catcode `\^12\catcode `\_12\catcode `\%12\relax}%
\providecommand \@@startlink[1]{}%
\providecommand \@@endlink[0]{}%
\providecommand \url  [0]{\begingroup\@sanitize@url \@url }%
\providecommand \@url [1]{\endgroup\@href {#1}{\urlprefix }}%
\providecommand \urlprefix  [0]{URL }%
\providecommand \Eprint [0]{\href }%
\providecommand \doibase [0]{http://dx.doi.org/}%
\providecommand \selectlanguage [0]{\@gobble}%
\providecommand \bibinfo  [0]{\@secondoftwo}%
\providecommand \bibfield  [0]{\@secondoftwo}%
\providecommand \translation [1]{[#1]}%
\providecommand \BibitemOpen [0]{}%
\providecommand \bibitemStop [0]{}%
\providecommand \bibitemNoStop [0]{.\EOS\space}%
\providecommand \EOS [0]{\spacefactor3000\relax}%
\providecommand \BibitemShut  [1]{\csname bibitem#1\endcsname}%
\let\auto@bib@innerbib\@empty
\bibitem [{\citenamefont {Slonczewski}(1997)}]{slonczewski1997}%
  \BibitemOpen
  \bibfield  {author} {\bibinfo {author} {\bibfnamefont {J.~C.}\ \bibnamefont
  {Slonczewski}},\ }\href {http://www.google.com/patents/US5695864} {\enquote
  {\bibinfo {title} {Electronic device using magnetic components},}\ }\bibinfo
  {howpublished} {{U.S. Patent No.} 5,695,864} (\bibinfo {year}
  {1997})\BibitemShut {NoStop}%
\bibitem [{\citenamefont {Kiselev}\ \emph {et~al.}(2003)\citenamefont
  {Kiselev}, \citenamefont {Sankey}, \citenamefont {Krivorotov}, \citenamefont
  {Emley}, \citenamefont {Schoelkopf}, \citenamefont {Buhrman},\ and\
  \citenamefont {Ralph}}]{kiselev2003}%
  \BibitemOpen
  \bibfield  {author} {\bibinfo {author} {\bibfnamefont {S.~I.}\ \bibnamefont
  {Kiselev}}, \bibinfo {author} {\bibfnamefont {J.~C.}\ \bibnamefont {Sankey}},
  \bibinfo {author} {\bibfnamefont {I.~N.}\ \bibnamefont {Krivorotov}},
  \bibinfo {author} {\bibfnamefont {N.~C.}\ \bibnamefont {Emley}}, \bibinfo
  {author} {\bibfnamefont {R.~J.}\ \bibnamefont {Schoelkopf}}, \bibinfo
  {author} {\bibfnamefont {R.~A.}\ \bibnamefont {Buhrman}}, \ and\ \bibinfo
  {author} {\bibfnamefont {D.~C.}\ \bibnamefont {Ralph}},\ }\href
  {http://dx.doi.org/10.1038/nature01967} {\bibfield  {journal} {\bibinfo
  {journal} {Nature}\ }\textbf {\bibinfo {volume} {425}},\ \bibinfo {pages}
  {380{\textendash}383} (\bibinfo {year} {2003})}\BibitemShut {NoStop}%
\bibitem [{\citenamefont {Tulapurkar}\ \emph {et~al.}(2005)\citenamefont
  {Tulapurkar}, \citenamefont {Suzuki}, \citenamefont {Fukushima},
  \citenamefont {Kubota}, \citenamefont {Maehara}, \citenamefont {Tsunekawa},
  \citenamefont {Djayaprawira}, \citenamefont {Watanabe},\ and\ \citenamefont
  {Yuasa}}]{tulapurkar2005}%
  \BibitemOpen
  \bibfield  {author} {\bibinfo {author} {\bibfnamefont {A.~A.}\ \bibnamefont
  {Tulapurkar}}, \bibinfo {author} {\bibfnamefont {Y.}~\bibnamefont {Suzuki}},
  \bibinfo {author} {\bibfnamefont {A.}~\bibnamefont {Fukushima}}, \bibinfo
  {author} {\bibfnamefont {H.}~\bibnamefont {Kubota}}, \bibinfo {author}
  {\bibfnamefont {H.}~\bibnamefont {Maehara}}, \bibinfo {author} {\bibfnamefont
  {K.}~\bibnamefont {Tsunekawa}}, \bibinfo {author} {\bibfnamefont {D.~D.}\
  \bibnamefont {Djayaprawira}}, \bibinfo {author} {\bibfnamefont
  {N.}~\bibnamefont {Watanabe}}, \ and\ \bibinfo {author} {\bibfnamefont
  {S.}~\bibnamefont {Yuasa}},\ }\href {\doibase 10.1038/nature04207} {\bibfield
   {journal} {\bibinfo  {journal} {Nature}\ }\textbf {\bibinfo {volume}
  {438}},\ \bibinfo {pages} {339} (\bibinfo {year} {2005})}\BibitemShut
  {NoStop}%
\bibitem [{\citenamefont {Miwa}\ \emph {et~al.}(2014)\citenamefont {Miwa},
  \citenamefont {Ishibashi}, \citenamefont {Tomita}, \citenamefont {Nozaki},
  \citenamefont {Tamura}, \citenamefont {Ando}, \citenamefont {Mizuochi},
  \citenamefont {Saruya}, \citenamefont {Kubota}, \citenamefont {Yakushiji},
  \citenamefont {Taniguchi}, \citenamefont {Imamura}, \citenamefont
  {Fukushima}, \citenamefont {Yuasa},\ and\ \citenamefont {Suzuki}}]{miwa2014}%
  \BibitemOpen
  \bibfield  {author} {\bibinfo {author} {\bibfnamefont {S.}~\bibnamefont
  {Miwa}}, \bibinfo {author} {\bibfnamefont {S.}~\bibnamefont {Ishibashi}},
  \bibinfo {author} {\bibfnamefont {H.}~\bibnamefont {Tomita}}, \bibinfo
  {author} {\bibfnamefont {T.}~\bibnamefont {Nozaki}}, \bibinfo {author}
  {\bibfnamefont {E.}~\bibnamefont {Tamura}}, \bibinfo {author} {\bibfnamefont
  {K.}~\bibnamefont {Ando}}, \bibinfo {author} {\bibfnamefont {N.}~\bibnamefont
  {Mizuochi}}, \bibinfo {author} {\bibfnamefont {T.}~\bibnamefont {Saruya}},
  \bibinfo {author} {\bibfnamefont {H.}~\bibnamefont {Kubota}}, \bibinfo
  {author} {\bibfnamefont {K.}~\bibnamefont {Yakushiji}}, \bibinfo {author}
  {\bibfnamefont {T.}~\bibnamefont {Taniguchi}}, \bibinfo {author}
  {\bibfnamefont {H.}~\bibnamefont {Imamura}}, \bibinfo {author} {\bibfnamefont
  {A.}~\bibnamefont {Fukushima}}, \bibinfo {author} {\bibfnamefont
  {S.}~\bibnamefont {Yuasa}}, \ and\ \bibinfo {author} {\bibfnamefont
  {Y.}~\bibnamefont {Suzuki}},\ }\href {\doibase 10.1038/nmat3778} {\bibfield
  {journal} {\bibinfo  {journal} {Nature Materials}\ }\textbf {\bibinfo
  {volume} {13}},\ \bibinfo {pages} {50} (\bibinfo {year} {2014})}\BibitemShut
  {NoStop}%
\bibitem [{\citenamefont {Fu}\ \emph {et~al.}(2012)\citenamefont {Fu},
  \citenamefont {Cao}, \citenamefont {Hemour}, \citenamefont {Wu},
  \citenamefont {Houssameddine}, \citenamefont {Lu}, \citenamefont {Pistorius},
  \citenamefont {Gui},\ and\ \citenamefont {Hu}}]{fu2012}%
  \BibitemOpen
  \bibfield  {author} {\bibinfo {author} {\bibfnamefont {L.}~\bibnamefont
  {Fu}}, \bibinfo {author} {\bibfnamefont {Z.~X.}\ \bibnamefont {Cao}},
  \bibinfo {author} {\bibfnamefont {S.}~\bibnamefont {Hemour}}, \bibinfo
  {author} {\bibfnamefont {K.}~\bibnamefont {Wu}}, \bibinfo {author}
  {\bibfnamefont {D.}~\bibnamefont {Houssameddine}}, \bibinfo {author}
  {\bibfnamefont {W.}~\bibnamefont {Lu}}, \bibinfo {author} {\bibfnamefont
  {S.}~\bibnamefont {Pistorius}}, \bibinfo {author} {\bibfnamefont {Y.~S.}\
  \bibnamefont {Gui}}, \ and\ \bibinfo {author} {\bibfnamefont
  {C.}~\bibnamefont {Hu}},\ }\href {\doibase 10.1063/1.4769837} {\bibfield
  {journal} {\bibinfo  {journal} {Applied Physics Letters}\ }\textbf {\bibinfo
  {volume} {101}},\ \bibinfo {pages} {232406} (\bibinfo {year}
  {2012})}\BibitemShut {NoStop}%
\bibitem [{\citenamefont {Fu}\ \emph {et~al.}(2014)\citenamefont {Fu},
  \citenamefont {Lu}, \citenamefont {Herrera}, \citenamefont {Tapia},
  \citenamefont {Gui}, \citenamefont {Pistorius},\ and\ \citenamefont
  {Hu}}]{fu2014}%
  \BibitemOpen
  \bibfield  {author} {\bibinfo {author} {\bibfnamefont {L.}~\bibnamefont
  {Fu}}, \bibinfo {author} {\bibfnamefont {W.}~\bibnamefont {Lu}}, \bibinfo
  {author} {\bibfnamefont {D.~R.}\ \bibnamefont {Herrera}}, \bibinfo {author}
  {\bibfnamefont {D.~F.}\ \bibnamefont {Tapia}}, \bibinfo {author}
  {\bibfnamefont {Y.~S.}\ \bibnamefont {Gui}}, \bibinfo {author} {\bibfnamefont
  {S.}~\bibnamefont {Pistorius}}, \ and\ \bibinfo {author} {\bibfnamefont
  {C.}~\bibnamefont {Hu}},\ }\href {\doibase 10.1063/1.4896691} {\bibfield
  {journal} {\bibinfo  {journal} {Applied Physics Letters}\ }\textbf {\bibinfo
  {volume} {105}},\ \bibinfo {pages} {122406} (\bibinfo {year}
  {2014})}\BibitemShut {NoStop}%
\bibitem [{\citenamefont {Kim}\ \emph {et~al.}(2008)\citenamefont {Kim},
  \citenamefont {Tiberkevich},\ and\ \citenamefont {Slavin}}]{kim2008}%
  \BibitemOpen
  \bibfield  {author} {\bibinfo {author} {\bibfnamefont {J.}~\bibnamefont
  {Kim}}, \bibinfo {author} {\bibfnamefont {V.}~\bibnamefont {Tiberkevich}}, \
  and\ \bibinfo {author} {\bibfnamefont {A.~N.}\ \bibnamefont {Slavin}},\
  }\href {\doibase 10.1103/PhysRevLett.100.017207} {\bibfield  {journal}
  {\bibinfo  {journal} {Physical Review Letters}\ }\textbf {\bibinfo {volume}
  {100}},\ \bibinfo {pages} {017207} (\bibinfo {year} {2008})}\BibitemShut
  {NoStop}%
\bibitem [{\citenamefont {Georges}\ \emph {et~al.}(2008)\citenamefont
  {Georges}, \citenamefont {Grollier}, \citenamefont {Darques}, \citenamefont
  {Cros}, \citenamefont {Deranlot}, \citenamefont {Marcilhac}, \citenamefont
  {Faini},\ and\ \citenamefont {Fert}}]{georges2008-a}%
  \BibitemOpen
  \bibfield  {author} {\bibinfo {author} {\bibfnamefont {B.}~\bibnamefont
  {Georges}}, \bibinfo {author} {\bibfnamefont {J.}~\bibnamefont {Grollier}},
  \bibinfo {author} {\bibfnamefont {M.}~\bibnamefont {Darques}}, \bibinfo
  {author} {\bibfnamefont {V.}~\bibnamefont {Cros}}, \bibinfo {author}
  {\bibfnamefont {C.}~\bibnamefont {Deranlot}}, \bibinfo {author}
  {\bibfnamefont {B.}~\bibnamefont {Marcilhac}}, \bibinfo {author}
  {\bibfnamefont {G.}~\bibnamefont {Faini}}, \ and\ \bibinfo {author}
  {\bibfnamefont {A.}~\bibnamefont {Fert}},\ }\href@noop {} {\bibfield
  {journal} {\bibinfo  {journal} {Physical Review Letters}\ }\textbf {\bibinfo
  {volume} {101}},\ \bibinfo {pages} {017201} (\bibinfo {year}
  {2008})}\BibitemShut {NoStop}%
\bibitem [{\citenamefont {Houssameddine}\ \emph {et~al.}(2008)\citenamefont
  {Houssameddine}, \citenamefont {Florez}, \citenamefont {Katine},
  \citenamefont {Michel}, \citenamefont {Ebels}, \citenamefont {Mauri},
  \citenamefont {Ozatay}, \citenamefont {Delaet}, \citenamefont {Viala},
  \citenamefont {Folks}, \citenamefont {Terris},\ and\ \citenamefont
  {Cyrille}}]{houssameddine2008}%
  \BibitemOpen
  \bibfield  {author} {\bibinfo {author} {\bibfnamefont {D.}~\bibnamefont
  {Houssameddine}}, \bibinfo {author} {\bibfnamefont {S.~H.}\ \bibnamefont
  {Florez}}, \bibinfo {author} {\bibfnamefont {J.~A.}\ \bibnamefont {Katine}},
  \bibinfo {author} {\bibfnamefont {J.}~\bibnamefont {Michel}}, \bibinfo
  {author} {\bibfnamefont {U.}~\bibnamefont {Ebels}}, \bibinfo {author}
  {\bibfnamefont {D.}~\bibnamefont {Mauri}}, \bibinfo {author} {\bibfnamefont
  {O.}~\bibnamefont {Ozatay}}, \bibinfo {author} {\bibfnamefont
  {B.}~\bibnamefont {Delaet}}, \bibinfo {author} {\bibfnamefont
  {B.}~\bibnamefont {Viala}}, \bibinfo {author} {\bibfnamefont
  {L.}~\bibnamefont {Folks}}, \bibinfo {author} {\bibfnamefont {B.~D.}\
  \bibnamefont {Terris}}, \ and\ \bibinfo {author} {\bibfnamefont
  {M.}~\bibnamefont {Cyrille}},\ }\href {\doibase
  http://dx.doi.org/10.1063/1.2956418} {\bibfield  {journal} {\bibinfo
  {journal} {Applied Physics Letters}\ }\textbf {\bibinfo {volume} {93}},\
  \bibinfo {pages} {022505} (\bibinfo {year} {2008})}\BibitemShut {NoStop}%
\bibitem [{\citenamefont {Maehara}(2014)}]{maehara2013}%
  \BibitemOpen
  \bibfield  {author} {\bibinfo {author} {\bibfnamefont {H.}~\bibnamefont
  {Maehara}},\ }\href {https://www.google.com/patents/US8836438} {\enquote
  {\bibinfo {title} {Oscillation element and method for manufacturing
  oscillation element},}\ }\bibinfo {howpublished} {U.S. Patent No. 8,836,438}
  (\bibinfo {year} {2014})\BibitemShut {NoStop}%
\bibitem [{\citenamefont {Maehara}\ \emph {et~al.}(2013)\citenamefont
  {Maehara}, \citenamefont {Kubota}, \citenamefont {Suzuki}, \citenamefont
  {Seki}, \citenamefont {Nishimura}, \citenamefont {Nagamine}, \citenamefont
  {Tsunekawa}, \citenamefont {Fukushima}, \citenamefont {Deac}, \citenamefont
  {Ando},\ and\ \citenamefont {Yuasa}}]{maehara2013-a}%
  \BibitemOpen
  \bibfield  {author} {\bibinfo {author} {\bibfnamefont {H.}~\bibnamefont
  {Maehara}}, \bibinfo {author} {\bibfnamefont {H.}~\bibnamefont {Kubota}},
  \bibinfo {author} {\bibfnamefont {Y.}~\bibnamefont {Suzuki}}, \bibinfo
  {author} {\bibfnamefont {T.}~\bibnamefont {Seki}}, \bibinfo {author}
  {\bibfnamefont {K.}~\bibnamefont {Nishimura}}, \bibinfo {author}
  {\bibfnamefont {Y.}~\bibnamefont {Nagamine}}, \bibinfo {author}
  {\bibfnamefont {K.}~\bibnamefont {Tsunekawa}}, \bibinfo {author}
  {\bibfnamefont {A.}~\bibnamefont {Fukushima}}, \bibinfo {author}
  {\bibfnamefont {A.~M.}\ \bibnamefont {Deac}}, \bibinfo {author}
  {\bibfnamefont {K.}~\bibnamefont {Ando}}, \ and\ \bibinfo {author}
  {\bibfnamefont {S.}~\bibnamefont {Yuasa}},\ }\href {\doibase
  10.7567/APEX.6.113005} {\bibfield  {journal} {\bibinfo  {journal} {Applied
  Physics Express}\ }\textbf {\bibinfo {volume} {6}},\ \bibinfo {pages}
  {113005} (\bibinfo {year} {2013})}\BibitemShut {NoStop}%
\bibitem [{\citenamefont {Maehara}\ \emph {et~al.}(2014)\citenamefont
  {Maehara}, \citenamefont {Kubota}, \citenamefont {Suzuki}, \citenamefont
  {Seki}, \citenamefont {Nishimura}, \citenamefont {Nagamine}, \citenamefont
  {Tsunekawa}, \citenamefont {Fukushima}, \citenamefont {Arai}, \citenamefont
  {Taniguchi}, \citenamefont {Imamura}, \citenamefont {Ando},\ and\
  \citenamefont {Yuasa}}]{maehara2014}%
  \BibitemOpen
  \bibfield  {author} {\bibinfo {author} {\bibfnamefont {H.}~\bibnamefont
  {Maehara}}, \bibinfo {author} {\bibfnamefont {H.}~\bibnamefont {Kubota}},
  \bibinfo {author} {\bibfnamefont {Y.}~\bibnamefont {Suzuki}}, \bibinfo
  {author} {\bibfnamefont {T.}~\bibnamefont {Seki}}, \bibinfo {author}
  {\bibfnamefont {K.}~\bibnamefont {Nishimura}}, \bibinfo {author}
  {\bibfnamefont {Y.}~\bibnamefont {Nagamine}}, \bibinfo {author}
  {\bibfnamefont {K.}~\bibnamefont {Tsunekawa}}, \bibinfo {author}
  {\bibfnamefont {A.}~\bibnamefont {Fukushima}}, \bibinfo {author}
  {\bibfnamefont {H.}~\bibnamefont {Arai}}, \bibinfo {author} {\bibfnamefont
  {T.}~\bibnamefont {Taniguchi}}, \bibinfo {author} {\bibfnamefont
  {H.}~\bibnamefont {Imamura}}, \bibinfo {author} {\bibfnamefont
  {K.}~\bibnamefont {Ando}}, \ and\ \bibinfo {author} {\bibfnamefont
  {S.}~\bibnamefont {Yuasa}},\ }\href {\doibase 10.7567/APEX.7.023003}
  {\bibfield  {journal} {\bibinfo  {journal} {Applied Physics Express}\
  }\textbf {\bibinfo {volume} {7}},\ \bibinfo {pages} {023003} (\bibinfo {year}
  {2014})}\BibitemShut {NoStop}%
\bibitem [{\citenamefont {Houssameddine}\ \emph {et~al.}(2009)\citenamefont
  {Houssameddine}, \citenamefont {Ebels}, \citenamefont {Dieny}, \citenamefont
  {Garello}, \citenamefont {Michel}, \citenamefont {Delaet}, \citenamefont
  {Viala}, \citenamefont {Cyrille}, \citenamefont {Katine},\ and\ \citenamefont
  {Mauri}}]{houssameddine2009}%
  \BibitemOpen
  \bibfield  {author} {\bibinfo {author} {\bibfnamefont {D.}~\bibnamefont
  {Houssameddine}}, \bibinfo {author} {\bibfnamefont {U.}~\bibnamefont
  {Ebels}}, \bibinfo {author} {\bibfnamefont {B.}~\bibnamefont {Dieny}},
  \bibinfo {author} {\bibfnamefont {K.}~\bibnamefont {Garello}}, \bibinfo
  {author} {\bibfnamefont {J.}~\bibnamefont {Michel}}, \bibinfo {author}
  {\bibfnamefont {B.}~\bibnamefont {Delaet}}, \bibinfo {author} {\bibfnamefont
  {B.}~\bibnamefont {Viala}}, \bibinfo {author} {\bibfnamefont
  {M.}~\bibnamefont {Cyrille}}, \bibinfo {author} {\bibfnamefont {J.~A.}\
  \bibnamefont {Katine}}, \ and\ \bibinfo {author} {\bibfnamefont
  {D.}~\bibnamefont {Mauri}},\ }\href {\doibase 10.1103/PhysRevLett.102.257202}
  {\bibfield  {journal} {\bibinfo  {journal} {Physical Review Letters}\
  }\textbf {\bibinfo {volume} {102}},\ \bibinfo {pages} {257202} (\bibinfo
  {year} {2009})}\BibitemShut {NoStop}%
\bibitem [{\citenamefont {Devolder}\ \emph {et~al.}(2009)\citenamefont
  {Devolder}, \citenamefont {Bianchini}, \citenamefont {Kim}, \citenamefont
  {Crozat}, \citenamefont {Chappert}, \citenamefont {Cornelissen},
  \citenamefont {Beeck},\ and\ \citenamefont {Lagae}}]{devolder2009}%
  \BibitemOpen
  \bibfield  {author} {\bibinfo {author} {\bibfnamefont {T.}~\bibnamefont
  {Devolder}}, \bibinfo {author} {\bibfnamefont {L.}~\bibnamefont {Bianchini}},
  \bibinfo {author} {\bibfnamefont {J.}~\bibnamefont {Kim}}, \bibinfo {author}
  {\bibfnamefont {P.}~\bibnamefont {Crozat}}, \bibinfo {author} {\bibfnamefont
  {C.}~\bibnamefont {Chappert}}, \bibinfo {author} {\bibfnamefont
  {S.}~\bibnamefont {Cornelissen}}, \bibinfo {author} {\bibfnamefont
  {M.~O.~d.}\ \bibnamefont {Beeck}}, \ and\ \bibinfo {author} {\bibfnamefont
  {L.}~\bibnamefont {Lagae}},\ }\href {\doibase 10.1063/1.3260233} {\bibfield
  {journal} {\bibinfo  {journal} {Journal of Applied Physics}\ }\textbf
  {\bibinfo {volume} {106}},\ \bibinfo {pages} {103921} (\bibinfo {year}
  {2009})}\BibitemShut {NoStop}%
\bibitem [{\citenamefont {Kudo}\ \emph {et~al.}(2009)\citenamefont {Kudo},
  \citenamefont {Nagasawa}, \citenamefont {Sato},\ and\ \citenamefont
  {Mizushima}}]{kudo2009}%
  \BibitemOpen
  \bibfield  {author} {\bibinfo {author} {\bibfnamefont {K.}~\bibnamefont
  {Kudo}}, \bibinfo {author} {\bibfnamefont {T.}~\bibnamefont {Nagasawa}},
  \bibinfo {author} {\bibfnamefont {R.}~\bibnamefont {Sato}}, \ and\ \bibinfo
  {author} {\bibfnamefont {K.}~\bibnamefont {Mizushima}},\ }\href {\doibase
  10.1063/1.3056407} {\bibfield  {journal} {\bibinfo  {journal} {Journal of
  Applied Physics}\ }\textbf {\bibinfo {volume} {105}},\ \bibinfo {pages}
  {07D105} (\bibinfo {year} {2009})}\BibitemShut {NoStop}%
\bibitem [{\citenamefont {Fukuzawa}\ \emph {et~al.}(2004)\citenamefont
  {Fukuzawa}, \citenamefont {Yuasa}, \citenamefont {Hashimoto}, \citenamefont
  {Koi}, \citenamefont {Iwasaki}, \citenamefont {Takagishi}, \citenamefont
  {Tanaka},\ and\ \citenamefont {Sahashi}}]{fukuzawa2004}%
  \BibitemOpen
  \bibfield  {author} {\bibinfo {author} {\bibfnamefont {H.}~\bibnamefont
  {Fukuzawa}}, \bibinfo {author} {\bibfnamefont {H.}~\bibnamefont {Yuasa}},
  \bibinfo {author} {\bibfnamefont {S.}~\bibnamefont {Hashimoto}}, \bibinfo
  {author} {\bibfnamefont {K.}~\bibnamefont {Koi}}, \bibinfo {author}
  {\bibfnamefont {H.}~\bibnamefont {Iwasaki}}, \bibinfo {author} {\bibfnamefont
  {M.}~\bibnamefont {Takagishi}}, \bibinfo {author} {\bibfnamefont
  {Y.}~\bibnamefont {Tanaka}}, \ and\ \bibinfo {author} {\bibfnamefont
  {M.}~\bibnamefont {Sahashi}},\ }\href {\doibase 10.1109/TMAG.2004.829185}
  {\bibfield  {journal} {\bibinfo  {journal} {{IEEE} Transactions on
  Magnetics}\ }\textbf {\bibinfo {volume} {40}},\ \bibinfo {pages} {2236}
  (\bibinfo {year} {2004})}\BibitemShut {NoStop}%
\bibitem [{\citenamefont {Fuke}\ \emph {et~al.}(2007)\citenamefont {Fuke},
  \citenamefont {Hashimoto}, \citenamefont {Takagishi}, \citenamefont
  {Iwasaki}, \citenamefont {Kawasaki}, \citenamefont {Miyake},\ and\
  \citenamefont {Sahashi}}]{fuke2007}%
  \BibitemOpen
  \bibfield  {author} {\bibinfo {author} {\bibfnamefont {H.~N.}\ \bibnamefont
  {Fuke}}, \bibinfo {author} {\bibfnamefont {S.}~\bibnamefont {Hashimoto}},
  \bibinfo {author} {\bibfnamefont {M.}~\bibnamefont {Takagishi}}, \bibinfo
  {author} {\bibfnamefont {H.}~\bibnamefont {Iwasaki}}, \bibinfo {author}
  {\bibfnamefont {S.}~\bibnamefont {Kawasaki}}, \bibinfo {author}
  {\bibfnamefont {K.}~\bibnamefont {Miyake}}, \ and\ \bibinfo {author}
  {\bibfnamefont {M.}~\bibnamefont {Sahashi}},\ }\href {\doibase
  10.1109/TMAG.2007.893117} {\bibfield  {journal} {\bibinfo  {journal} {{IEEE}
  Transactions on Magnetics}\ }\textbf {\bibinfo {volume} {43}},\ \bibinfo
  {pages} {2848} (\bibinfo {year} {2007})}\BibitemShut {NoStop}%
\bibitem [{\citenamefont {Takagishi}\ \emph {et~al.}(2009)\citenamefont
  {Takagishi}, \citenamefont {Fuke}, \citenamefont {Hashimoto}, \citenamefont
  {Iwasaki}, \citenamefont {Kawasaki}, \citenamefont {Shiozaki},\ and\
  \citenamefont {Sahashi}}]{takagishi2009}%
  \BibitemOpen
  \bibfield  {author} {\bibinfo {author} {\bibfnamefont {M.}~\bibnamefont
  {Takagishi}}, \bibinfo {author} {\bibfnamefont {H.~N.}\ \bibnamefont {Fuke}},
  \bibinfo {author} {\bibfnamefont {S.}~\bibnamefont {Hashimoto}}, \bibinfo
  {author} {\bibfnamefont {H.}~\bibnamefont {Iwasaki}}, \bibinfo {author}
  {\bibfnamefont {S.}~\bibnamefont {Kawasaki}}, \bibinfo {author}
  {\bibfnamefont {R.}~\bibnamefont {Shiozaki}}, \ and\ \bibinfo {author}
  {\bibfnamefont {M.}~\bibnamefont {Sahashi}},\ }\href@noop {} {\bibfield
  {journal} {\bibinfo  {journal} {Journal of Applied Physics}\ }\textbf
  {\bibinfo {volume} {105}},\ \bibinfo {pages} {07B725} (\bibinfo {year}
  {2009})}\BibitemShut {NoStop}%
\bibitem [{\citenamefont {Shiokawa}\ \emph {et~al.}(2011)\citenamefont
  {Shiokawa}, \citenamefont {Shiota}, \citenamefont {Watanabe}, \citenamefont
  {Otsuka}, \citenamefont {Doi},\ and\ \citenamefont {Sahashi}}]{shiokawa2011}%
  \BibitemOpen
  \bibfield  {author} {\bibinfo {author} {\bibfnamefont {Y.}~\bibnamefont
  {Shiokawa}}, \bibinfo {author} {\bibfnamefont {M.}~\bibnamefont {Shiota}},
  \bibinfo {author} {\bibfnamefont {Y.}~\bibnamefont {Watanabe}}, \bibinfo
  {author} {\bibfnamefont {T.}~\bibnamefont {Otsuka}}, \bibinfo {author}
  {\bibfnamefont {M.}~\bibnamefont {Doi}}, \ and\ \bibinfo {author}
  {\bibfnamefont {M.}~\bibnamefont {Sahashi}},\ }\href {\doibase
  10.1109/TMAG.2011.2157110} {\bibfield  {journal} {\bibinfo  {journal}
  {Magnetics, {IEEE} Transactions on}\ }\textbf {\bibinfo {volume} {47}},\
  \bibinfo {pages} {3470 } (\bibinfo {year} {2011})}\BibitemShut {NoStop}%
\bibitem [{\citenamefont {Yuasa}\ \emph {et~al.}(2013)\citenamefont {Yuasa},
  \citenamefont {Hara}, \citenamefont {Fuji},\ and\ \citenamefont
  {Fukuzawa}}]{yuasa2013}%
  \BibitemOpen
  \bibfield  {author} {\bibinfo {author} {\bibfnamefont {H.}~\bibnamefont
  {Yuasa}}, \bibinfo {author} {\bibfnamefont {M.}~\bibnamefont {Hara}},
  \bibinfo {author} {\bibfnamefont {Y.}~\bibnamefont {Fuji}}, \ and\ \bibinfo
  {author} {\bibfnamefont {H.}~\bibnamefont {Fukuzawa}},\ }\href {\doibase
  10.1209/0295-5075/101/47005} {\bibfield  {journal} {\bibinfo  {journal}
  {{EPL} {(Europhysics} Letters)}\ }\textbf {\bibinfo {volume} {101}},\
  \bibinfo {pages} {47005} (\bibinfo {year} {2013})}\BibitemShut {NoStop}%
\bibitem [{\citenamefont {Endo}\ \emph {et~al.}(2009)\citenamefont {Endo},
  \citenamefont {Tanaka}, \citenamefont {Doi}, \citenamefont {Hashimoto},
  \citenamefont {Fuke}, \citenamefont {Iwasaki},\ and\ \citenamefont
  {Sahashi}}]{endo2009}%
  \BibitemOpen
  \bibfield  {author} {\bibinfo {author} {\bibfnamefont {H.}~\bibnamefont
  {Endo}}, \bibinfo {author} {\bibfnamefont {T.}~\bibnamefont {Tanaka}},
  \bibinfo {author} {\bibfnamefont {M.}~\bibnamefont {Doi}}, \bibinfo {author}
  {\bibfnamefont {S.}~\bibnamefont {Hashimoto}}, \bibinfo {author}
  {\bibfnamefont {H.~N.}\ \bibnamefont {Fuke}}, \bibinfo {author}
  {\bibfnamefont {H.}~\bibnamefont {Iwasaki}}, \ and\ \bibinfo {author}
  {\bibfnamefont {M.}~\bibnamefont {Sahashi}},\ }\href {\doibase
  10.1109/TMAG.2009.2023851} {\bibfield  {journal} {\bibinfo  {journal} {{IEEE}
  Transactions on Magnetics}\ }\textbf {\bibinfo {volume} {45}},\ \bibinfo
  {pages} {3418} (\bibinfo {year} {2009})}\BibitemShut {NoStop}%
\bibitem [{\citenamefont {Suzuki}\ \emph {et~al.}(2009)\citenamefont {Suzuki},
  \citenamefont {Endo}, \citenamefont {Nakamura}, \citenamefont {Tanaka},
  \citenamefont {Doi}, \citenamefont {Hashimoto}, \citenamefont {Fuke},
  \citenamefont {Takagishi}, \citenamefont {Iwasaki},\ and\ \citenamefont
  {Sahashi}}]{suzuki2009}%
  \BibitemOpen
  \bibfield  {author} {\bibinfo {author} {\bibfnamefont {H.}~\bibnamefont
  {Suzuki}}, \bibinfo {author} {\bibfnamefont {H.}~\bibnamefont {Endo}},
  \bibinfo {author} {\bibfnamefont {T.}~\bibnamefont {Nakamura}}, \bibinfo
  {author} {\bibfnamefont {T.}~\bibnamefont {Tanaka}}, \bibinfo {author}
  {\bibfnamefont {M.}~\bibnamefont {Doi}}, \bibinfo {author} {\bibfnamefont
  {S.}~\bibnamefont {Hashimoto}}, \bibinfo {author} {\bibfnamefont {H.~N.}\
  \bibnamefont {Fuke}}, \bibinfo {author} {\bibfnamefont {M.}~\bibnamefont
  {Takagishi}}, \bibinfo {author} {\bibfnamefont {H.}~\bibnamefont {Iwasaki}},
  \ and\ \bibinfo {author} {\bibfnamefont {M.}~\bibnamefont {Sahashi}},\ }\href
  {\doibase doi:10.1063/1.3076047} {\bibfield  {journal} {\bibinfo  {journal}
  {Journal of Applied Physics}\ }\textbf {\bibinfo {volume} {105}},\ \bibinfo
  {pages} {07D124} (\bibinfo {year} {2009})}\BibitemShut {NoStop}%
\bibitem [{\citenamefont {Suzuki}\ \emph {et~al.}(2011)\citenamefont {Suzuki},
  \citenamefont {Nakamura}, \citenamefont {Endo}, \citenamefont {Doi},
  \citenamefont {Tsukahara}, \citenamefont {Imamura}, \citenamefont {Fuke},
  \citenamefont {Hashimoto}, \citenamefont {Iwasaki},\ and\ \citenamefont
  {Sahashi}}]{suzuki2011}%
  \BibitemOpen
  \bibfield  {author} {\bibinfo {author} {\bibfnamefont {H.}~\bibnamefont
  {Suzuki}}, \bibinfo {author} {\bibfnamefont {T.}~\bibnamefont {Nakamura}},
  \bibinfo {author} {\bibfnamefont {H.}~\bibnamefont {Endo}}, \bibinfo {author}
  {\bibfnamefont {M.}~\bibnamefont {Doi}}, \bibinfo {author} {\bibfnamefont
  {H.}~\bibnamefont {Tsukahara}}, \bibinfo {author} {\bibfnamefont
  {H.}~\bibnamefont {Imamura}}, \bibinfo {author} {\bibfnamefont {H.~N.}\
  \bibnamefont {Fuke}}, \bibinfo {author} {\bibfnamefont {S.}~\bibnamefont
  {Hashimoto}}, \bibinfo {author} {\bibfnamefont {H.}~\bibnamefont {Iwasaki}},
  \ and\ \bibinfo {author} {\bibfnamefont {M.}~\bibnamefont {Sahashi}},\ }\href
  {\doibase doi:10.1063/1.3619835} {\bibfield  {journal} {\bibinfo  {journal}
  {Applied Physics Letters}\ }\textbf {\bibinfo {volume} {99}},\ \bibinfo
  {pages} {092507} (\bibinfo {year} {2011})}\BibitemShut {NoStop}%
\bibitem [{\citenamefont {Doi}\ \emph {et~al.}(2011)\citenamefont {Doi},
  \citenamefont {Endo}, \citenamefont {Shirafuji}, \citenamefont {Kawasaki},
  \citenamefont {Sahashi}, \citenamefont {Fuke}, \citenamefont {Iwasaki},\ and\
  \citenamefont {Imamura}}]{doi2011}%
  \BibitemOpen
  \bibfield  {author} {\bibinfo {author} {\bibfnamefont {M.}~\bibnamefont
  {Doi}}, \bibinfo {author} {\bibfnamefont {H.}~\bibnamefont {Endo}}, \bibinfo
  {author} {\bibfnamefont {K.}~\bibnamefont {Shirafuji}}, \bibinfo {author}
  {\bibfnamefont {S.}~\bibnamefont {Kawasaki}}, \bibinfo {author}
  {\bibfnamefont {M.}~\bibnamefont {Sahashi}}, \bibinfo {author} {\bibfnamefont
  {H.~N.}\ \bibnamefont {Fuke}}, \bibinfo {author} {\bibfnamefont
  {H.}~\bibnamefont {Iwasaki}}, \ and\ \bibinfo {author} {\bibfnamefont
  {H.}~\bibnamefont {Imamura}},\ }\href
  {http://stacks.iop.org/0022-3727/44/i=9/a=092001} {\bibfield  {journal}
  {\bibinfo  {journal} {Journal of Physics D: Applied Physics}\ }\textbf
  {\bibinfo {volume} {44}},\ \bibinfo {pages} {092001} (\bibinfo {year}
  {2011})}\BibitemShut {NoStop}%
\bibitem [{\citenamefont {{Al-Mahdawi}}\ \emph {et~al.}(2011)\citenamefont
  {{Al-Mahdawi}}, \citenamefont {Doi}, \citenamefont {Hashimoto}, \citenamefont
  {Fuke}, \citenamefont {Iwasaki},\ and\ \citenamefont
  {Sahashi}}]{al-mahdawi2011}%
  \BibitemOpen
  \bibfield  {author} {\bibinfo {author} {\bibfnamefont {M.}~\bibnamefont
  {{Al-Mahdawi}}}, \bibinfo {author} {\bibfnamefont {M.}~\bibnamefont {Doi}},
  \bibinfo {author} {\bibfnamefont {S.}~\bibnamefont {Hashimoto}}, \bibinfo
  {author} {\bibfnamefont {H.~N.}\ \bibnamefont {Fuke}}, \bibinfo {author}
  {\bibfnamefont {H.}~\bibnamefont {Iwasaki}}, \ and\ \bibinfo {author}
  {\bibfnamefont {M.}~\bibnamefont {Sahashi}},\ }\href {\doibase
  10.1109/TMAG.2011.2159104} {\bibfield  {journal} {\bibinfo  {journal} {{IEEE}
  Transactions on Magnetics}\ }\textbf {\bibinfo {volume} {47}},\ \bibinfo
  {pages} {3380} (\bibinfo {year} {2011})}\BibitemShut {NoStop}%
\bibitem [{\citenamefont {{Al-Mahdawi}}\ and\ \citenamefont
  {Sahashi}(2014)}]{al-mahdawi2014}%
  \BibitemOpen
  \bibfield  {author} {\bibinfo {author} {\bibfnamefont {M.}~\bibnamefont
  {{Al-Mahdawi}}}\ and\ \bibinfo {author} {\bibfnamefont {M.}~\bibnamefont
  {Sahashi}},\ }\href {\doibase 10.1063/1.4862462} {\bibfield  {journal}
  {\bibinfo  {journal} {Applied Physics Letters}\ }\textbf {\bibinfo {volume}
  {104}},\ \bibinfo {pages} {032405} (\bibinfo {year} {2014})}\BibitemShut
  {NoStop}%
\bibitem [{\citenamefont {Tagirov}\ \emph {et~al.}(2001)\citenamefont
  {Tagirov}, \citenamefont {Vodopyanov},\ and\ \citenamefont
  {Efetov}}]{tagirov2001}%
  \BibitemOpen
  \bibfield  {author} {\bibinfo {author} {\bibfnamefont {L.~R.}\ \bibnamefont
  {Tagirov}}, \bibinfo {author} {\bibfnamefont {B.~P.}\ \bibnamefont
  {Vodopyanov}}, \ and\ \bibinfo {author} {\bibfnamefont {K.~B.}\ \bibnamefont
  {Efetov}},\ }\href {\doibase 10.1103/PhysRevB.63.104428} {\bibfield
  {journal} {\bibinfo  {journal} {Physical Review B}\ }\textbf {\bibinfo
  {volume} {63}},\ \bibinfo {pages} {104428} (\bibinfo {year}
  {2001})}\BibitemShut {NoStop}%
\bibitem [{\citenamefont {Sato}\ \emph {et~al.}(2009)\citenamefont {Sato},
  \citenamefont {Matsushita},\ and\ \citenamefont {Imamura}}]{sato2009}%
  \BibitemOpen
  \bibfield  {author} {\bibinfo {author} {\bibfnamefont {J.}~\bibnamefont
  {Sato}}, \bibinfo {author} {\bibfnamefont {K.}~\bibnamefont {Matsushita}}, \
  and\ \bibinfo {author} {\bibfnamefont {H.}~\bibnamefont {Imamura}},\ }\href
  {\doibase 10.1063/1.3055357} {\bibfield  {journal} {\bibinfo  {journal}
  {Journal of Applied Physics}\ }\textbf {\bibinfo {volume} {105}},\ \bibinfo
  {pages} {07D101 } (\bibinfo {year} {2009})}\BibitemShut {NoStop}%
\bibitem [{\citenamefont {Imamura}\ and\ \citenamefont
  {Sato}(2011)}]{imamura2011}%
  \BibitemOpen
  \bibfield  {author} {\bibinfo {author} {\bibfnamefont {H.}~\bibnamefont
  {Imamura}}\ and\ \bibinfo {author} {\bibfnamefont {J.}~\bibnamefont {Sato}},\
  }\href {\doibase 10.1088/1742-6596/266/1/012090} {\bibfield  {journal}
  {\bibinfo  {journal} {Journal of Physics: Conference Series}\ }\textbf
  {\bibinfo {volume} {266}},\ \bibinfo {pages} {012090} (\bibinfo {year}
  {2011})}\BibitemShut {NoStop}%
\bibitem [{\citenamefont {Shiokawa}\ \emph {et~al.}(2015)\citenamefont
  {Shiokawa}, \citenamefont {Jung}, \citenamefont {Otsuka},\ and\ \citenamefont
  {Sahashi}}]{shiokawa2015}%
  \BibitemOpen
  \bibfield  {author} {\bibinfo {author} {\bibfnamefont {Y.}~\bibnamefont
  {Shiokawa}}, \bibinfo {author} {\bibfnamefont {J.}~\bibnamefont {Jung}},
  \bibinfo {author} {\bibfnamefont {T.}~\bibnamefont {Otsuka}}, \ and\ \bibinfo
  {author} {\bibfnamefont {M.}~\bibnamefont {Sahashi}},\ }\href {\doibase
  10.1063/1.4927842} {\bibfield  {journal} {\bibinfo  {journal} {Journal of
  Applied Physics}\ }\textbf {\bibinfo {volume} {118}},\ \bibinfo {pages}
  {053909} (\bibinfo {year} {2015})}\BibitemShut {NoStop}%
\bibitem [{\citenamefont {Gr\"unberg}(1980)}]{grunberg1980}%
  \BibitemOpen
  \bibfield  {author} {\bibinfo {author} {\bibfnamefont {P.}~\bibnamefont
  {Gr\"unberg}},\ }\href {\doibase 10.1063/1.328292} {\bibfield  {journal}
  {\bibinfo  {journal} {Journal of Applied Physics}\ }\textbf {\bibinfo
  {volume} {51}},\ \bibinfo {pages} {4338} (\bibinfo {year}
  {1980})}\BibitemShut {NoStop}%
\bibitem [{\citenamefont {Seki}\ \emph {et~al.}(2010)\citenamefont {Seki},
  \citenamefont {Tomita}, \citenamefont {Shinjo},\ and\ \citenamefont
  {Suzuki}}]{seki2010}%
  \BibitemOpen
  \bibfield  {author} {\bibinfo {author} {\bibfnamefont {T.}~\bibnamefont
  {Seki}}, \bibinfo {author} {\bibfnamefont {H.}~\bibnamefont {Tomita}},
  \bibinfo {author} {\bibfnamefont {T.}~\bibnamefont {Shinjo}}, \ and\ \bibinfo
  {author} {\bibfnamefont {Y.}~\bibnamefont {Suzuki}},\ }\href {\doibase
  10.1063/1.3505357} {\bibfield  {journal} {\bibinfo  {journal} {Applied
  Physics Letters}\ }\textbf {\bibinfo {volume} {97}},\ \bibinfo {pages}
  {162508} (\bibinfo {year} {2010})}\BibitemShut {NoStop}%
\bibitem [{\citenamefont {Moriyama}\ \emph {et~al.}(2012)\citenamefont
  {Moriyama}, \citenamefont {Finocchio}, \citenamefont {Carpentieri},
  \citenamefont {Azzerboni}, \citenamefont {Ralph},\ and\ \citenamefont
  {Buhrman}}]{moriyama2012}%
  \BibitemOpen
  \bibfield  {author} {\bibinfo {author} {\bibfnamefont {T.}~\bibnamefont
  {Moriyama}}, \bibinfo {author} {\bibfnamefont {G.}~\bibnamefont {Finocchio}},
  \bibinfo {author} {\bibfnamefont {M.}~\bibnamefont {Carpentieri}}, \bibinfo
  {author} {\bibfnamefont {B.}~\bibnamefont {Azzerboni}}, \bibinfo {author}
  {\bibfnamefont {D.~C.}\ \bibnamefont {Ralph}}, \ and\ \bibinfo {author}
  {\bibfnamefont {R.~A.}\ \bibnamefont {Buhrman}},\ }\href {\doibase
  10.1103/PhysRevB.86.060411} {\bibfield  {journal} {\bibinfo  {journal}
  {Physical Review B}\ }\textbf {\bibinfo {volume} {86}},\ \bibinfo {pages}
  {060411} (\bibinfo {year} {2012})}\BibitemShut {NoStop}%
\bibitem [{\citenamefont {Braganca}\ \emph {et~al.}(2013)\citenamefont
  {Braganca}, \citenamefont {Pi}, \citenamefont {Zakai}, \citenamefont
  {Childress},\ and\ \citenamefont {Gurney}}]{braganca2013}%
  \BibitemOpen
  \bibfield  {author} {\bibinfo {author} {\bibfnamefont {P.~M.}\ \bibnamefont
  {Braganca}}, \bibinfo {author} {\bibfnamefont {K.}~\bibnamefont {Pi}},
  \bibinfo {author} {\bibfnamefont {R.}~\bibnamefont {Zakai}}, \bibinfo
  {author} {\bibfnamefont {J.~R.}\ \bibnamefont {Childress}}, \ and\ \bibinfo
  {author} {\bibfnamefont {B.~A.}\ \bibnamefont {Gurney}},\ }\href {\doibase
  10.1063/1.4838655} {\bibfield  {journal} {\bibinfo  {journal} {Applied
  Physics Letters}\ }\textbf {\bibinfo {volume} {103}},\ \bibinfo {pages}
  {232407} (\bibinfo {year} {2013})}\BibitemShut {NoStop}%
\bibitem [{\citenamefont {Nagasawa}\ \emph {et~al.}(2014)\citenamefont
  {Nagasawa}, \citenamefont {Kudo}, \citenamefont {Suto}, \citenamefont
  {Mizushima},\ and\ \citenamefont {Sato}}]{nagasawa2014}%
  \BibitemOpen
  \bibfield  {author} {\bibinfo {author} {\bibfnamefont {T.}~\bibnamefont
  {Nagasawa}}, \bibinfo {author} {\bibfnamefont {K.}~\bibnamefont {Kudo}},
  \bibinfo {author} {\bibfnamefont {H.}~\bibnamefont {Suto}}, \bibinfo {author}
  {\bibfnamefont {K.}~\bibnamefont {Mizushima}}, \ and\ \bibinfo {author}
  {\bibfnamefont {R.}~\bibnamefont {Sato}},\ }\href {\doibase
  10.1063/1.4901077} {\bibfield  {journal} {\bibinfo  {journal} {Applied
  Physics Letters}\ }\textbf {\bibinfo {volume} {105}},\ \bibinfo {pages}
  {182406} (\bibinfo {year} {2014})}\BibitemShut {NoStop}%
\bibitem [{\citenamefont {Fischbacher}\ \emph {et~al.}(2007)\citenamefont
  {Fischbacher}, \citenamefont {Franchin}, \citenamefont {Bordignon},\ and\
  \citenamefont {Fangohr}}]{fischbacher2007}%
  \BibitemOpen
  \bibfield  {author} {\bibinfo {author} {\bibfnamefont {T.}~\bibnamefont
  {Fischbacher}}, \bibinfo {author} {\bibfnamefont {M.}~\bibnamefont
  {Franchin}}, \bibinfo {author} {\bibfnamefont {G.}~\bibnamefont {Bordignon}},
  \ and\ \bibinfo {author} {\bibfnamefont {H.}~\bibnamefont {Fangohr}},\ }\href
  {\doibase 10.1109/TMAG.2007.893843} {\bibfield  {journal} {\bibinfo
  {journal} {Magnetics, {IEEE} Transactions on}\ }\textbf {\bibinfo {volume}
  {43}},\ \bibinfo {pages} {2896 } (\bibinfo {year} {2007})}\BibitemShut
  {NoStop}%
\bibitem [{\citenamefont {Miyake}\ \emph {et~al.}(2013)\citenamefont {Miyake},
  \citenamefont {Okutomi}, \citenamefont {Tsukahara}, \citenamefont {Imamura},\
  and\ \citenamefont {Sahashi}}]{miyake2013}%
  \BibitemOpen
  \bibfield  {author} {\bibinfo {author} {\bibfnamefont {K.}~\bibnamefont
  {Miyake}}, \bibinfo {author} {\bibfnamefont {Y.}~\bibnamefont {Okutomi}},
  \bibinfo {author} {\bibfnamefont {H.}~\bibnamefont {Tsukahara}}, \bibinfo
  {author} {\bibfnamefont {H.}~\bibnamefont {Imamura}}, \ and\ \bibinfo
  {author} {\bibfnamefont {M.}~\bibnamefont {Sahashi}},\ }\href {\doibase
  10.7567/APEX.6.113001} {\bibfield  {journal} {\bibinfo  {journal} {Applied
  Physics Express}\ }\textbf {\bibinfo {volume} {6}},\ \bibinfo {pages}
  {113001} (\bibinfo {year} {2013})}\BibitemShut {NoStop}%
\bibitem [{\citenamefont {Naidyuk}\ and\ \citenamefont
  {Yanson}(2005)}]{naidyuk2005}%
  \BibitemOpen
  \bibfield  {author} {\bibinfo {author} {\bibfnamefont {Y.~G.}\ \bibnamefont
  {Naidyuk}}\ and\ \bibinfo {author} {\bibfnamefont {I.~K.}\ \bibnamefont
  {Yanson}},\ }\href@noop {} {\emph {\bibinfo {title} {{Point-Contact}
  Spectroscopy}}}\ (\bibinfo  {publisher} {Springer},\ \bibinfo {year}
  {2005})\BibitemShut {NoStop}%
\bibitem [{\citenamefont {Strelkov}\ \emph {et~al.}(2011)\citenamefont
  {Strelkov}, \citenamefont {Vedyayev}, \citenamefont {Ryzhanova},
  \citenamefont {Gusakova}, \citenamefont {{Buda-Prejbeanu}}, \citenamefont
  {Chshiev}, \citenamefont {Amara}, \citenamefont {de~Mestier}, \citenamefont
  {Baraduc},\ and\ \citenamefont {Dieny}}]{strelkov2011}%
  \BibitemOpen
  \bibfield  {author} {\bibinfo {author} {\bibfnamefont {N.}~\bibnamefont
  {Strelkov}}, \bibinfo {author} {\bibfnamefont {A.}~\bibnamefont {Vedyayev}},
  \bibinfo {author} {\bibfnamefont {N.}~\bibnamefont {Ryzhanova}}, \bibinfo
  {author} {\bibfnamefont {D.}~\bibnamefont {Gusakova}}, \bibinfo {author}
  {\bibfnamefont {L.~D.}\ \bibnamefont {{Buda-Prejbeanu}}}, \bibinfo {author}
  {\bibfnamefont {M.}~\bibnamefont {Chshiev}}, \bibinfo {author} {\bibfnamefont
  {S.}~\bibnamefont {Amara}}, \bibinfo {author} {\bibfnamefont
  {N.}~\bibnamefont {de~Mestier}}, \bibinfo {author} {\bibfnamefont
  {C.}~\bibnamefont {Baraduc}}, \ and\ \bibinfo {author} {\bibfnamefont
  {B.}~\bibnamefont {Dieny}},\ }\href {\doibase 10.1103/PhysRevB.84.024416}
  {\bibfield  {journal} {\bibinfo  {journal} {Physical Review B}\ }\textbf
  {\bibinfo {volume} {84}},\ \bibinfo {pages} {024416} (\bibinfo {year}
  {2011})}\BibitemShut {NoStop}%
\bibitem [{\citenamefont {Dmytriiev}\ \emph {et~al.}(2010)\citenamefont
  {Dmytriiev}, \citenamefont {Meitzler}, \citenamefont {Bankowski},
  \citenamefont {Slavin},\ and\ \citenamefont {Tiberkevich}}]{dmytriiev2010}%
  \BibitemOpen
  \bibfield  {author} {\bibinfo {author} {\bibfnamefont {O.}~\bibnamefont
  {Dmytriiev}}, \bibinfo {author} {\bibfnamefont {T.}~\bibnamefont {Meitzler}},
  \bibinfo {author} {\bibfnamefont {E.}~\bibnamefont {Bankowski}}, \bibinfo
  {author} {\bibfnamefont {A.}~\bibnamefont {Slavin}}, \ and\ \bibinfo {author}
  {\bibfnamefont {V.}~\bibnamefont {Tiberkevich}},\ }\href {\doibase
  10.1088/0953-8984/22/13/136001} {\bibfield  {journal} {\bibinfo  {journal}
  {Journal of Physics: Condensed Matter}\ }\textbf {\bibinfo {volume} {22}},\
  \bibinfo {pages} {136001} (\bibinfo {year} {2010})}\BibitemShut {NoStop}%
\bibitem [{\citenamefont {Grimsditch}\ \emph {et~al.}(1996)\citenamefont
  {Grimsditch}, \citenamefont {Kumar},\ and\ \citenamefont
  {Fullerton}}]{grimsditch1996}%
  \BibitemOpen
  \bibfield  {author} {\bibinfo {author} {\bibfnamefont {M.}~\bibnamefont
  {Grimsditch}}, \bibinfo {author} {\bibfnamefont {S.}~\bibnamefont {Kumar}}, \
  and\ \bibinfo {author} {\bibfnamefont {E.~E.}\ \bibnamefont {Fullerton}},\
  }\href {\doibase 10.1103/PhysRevB.54.3385} {\bibfield  {journal} {\bibinfo
  {journal} {Physical Review B}\ }\textbf {\bibinfo {volume} {54}},\ \bibinfo
  {pages} {3385} (\bibinfo {year} {1996})}\BibitemShut {NoStop}%
\bibitem [{\citenamefont {Slavin}\ and\ \citenamefont
  {Tiberkevich}(2009)}]{slavin2009}%
  \BibitemOpen
  \bibfield  {author} {\bibinfo {author} {\bibfnamefont {A.}~\bibnamefont
  {Slavin}}\ and\ \bibinfo {author} {\bibfnamefont {V.}~\bibnamefont
  {Tiberkevich}},\ }\href {\doibase 10.1109/TMAG.2008.2009935} {\bibfield
  {journal} {\bibinfo  {journal} {Magnetics, {IEEE} Transactions on}\ }\textbf
  {\bibinfo {volume} {45}},\ \bibinfo {pages} {1875} (\bibinfo {year}
  {2009})}\BibitemShut {NoStop}%
\bibitem [{\citenamefont {Deac}\ \emph {et~al.}(2008)\citenamefont {Deac},
  \citenamefont {Fukushima}, \citenamefont {Kubota}, \citenamefont {Maehara},
  \citenamefont {Suzuki}, \citenamefont {Yuasa}, \citenamefont {Nagamine},
  \citenamefont {Tsunekawa}, \citenamefont {Djayaprawira},\ and\ \citenamefont
  {Watanabe}}]{deac2008}%
  \BibitemOpen
  \bibfield  {author} {\bibinfo {author} {\bibfnamefont {A.~M.}\ \bibnamefont
  {Deac}}, \bibinfo {author} {\bibfnamefont {A.}~\bibnamefont {Fukushima}},
  \bibinfo {author} {\bibfnamefont {H.}~\bibnamefont {Kubota}}, \bibinfo
  {author} {\bibfnamefont {H.}~\bibnamefont {Maehara}}, \bibinfo {author}
  {\bibfnamefont {Y.}~\bibnamefont {Suzuki}}, \bibinfo {author} {\bibfnamefont
  {S.}~\bibnamefont {Yuasa}}, \bibinfo {author} {\bibfnamefont
  {Y.}~\bibnamefont {Nagamine}}, \bibinfo {author} {\bibfnamefont
  {K.}~\bibnamefont {Tsunekawa}}, \bibinfo {author} {\bibfnamefont {D.~D.}\
  \bibnamefont {Djayaprawira}}, \ and\ \bibinfo {author} {\bibfnamefont
  {N.}~\bibnamefont {Watanabe}},\ }\href {\doibase 10.1038/nphys1036}
  {\bibfield  {journal} {\bibinfo  {journal} {Nature Physics}\ }\textbf
  {\bibinfo {volume} {4}},\ \bibinfo {pages} {803} (\bibinfo {year}
  {2008})}\BibitemShut {NoStop}%
\bibitem [{\citenamefont {Cochran}\ \emph {et~al.}(1990)\citenamefont
  {Cochran}, \citenamefont {Rudd}, \citenamefont {Muir}, \citenamefont
  {Heinrich},\ and\ \citenamefont {Celinski}}]{cochran1990}%
  \BibitemOpen
  \bibfield  {author} {\bibinfo {author} {\bibfnamefont {J.~F.}\ \bibnamefont
  {Cochran}}, \bibinfo {author} {\bibfnamefont {J.}~\bibnamefont {Rudd}},
  \bibinfo {author} {\bibfnamefont {W.~B.}\ \bibnamefont {Muir}}, \bibinfo
  {author} {\bibfnamefont {B.}~\bibnamefont {Heinrich}}, \ and\ \bibinfo
  {author} {\bibfnamefont {Z.}~\bibnamefont {Celinski}},\ }\href {\doibase
  10.1103/PhysRevB.42.508} {\bibfield  {journal} {\bibinfo  {journal} {Physical
  Review B}\ }\textbf {\bibinfo {volume} {42}},\ \bibinfo {pages} {508}
  (\bibinfo {year} {1990})}\BibitemShut {NoStop}%
\bibitem [{\citenamefont {Zivieri}\ \emph {et~al.}(2000)\citenamefont
  {Zivieri}, \citenamefont {Giovannini},\ and\ \citenamefont
  {Nizzoli}}]{zivieri2000}%
  \BibitemOpen
  \bibfield  {author} {\bibinfo {author} {\bibfnamefont {R.}~\bibnamefont
  {Zivieri}}, \bibinfo {author} {\bibfnamefont {L.}~\bibnamefont {Giovannini}},
  \ and\ \bibinfo {author} {\bibfnamefont {F.}~\bibnamefont {Nizzoli}},\ }\href
  {\doibase 10.1103/PhysRevB.62.14950} {\bibfield  {journal} {\bibinfo
  {journal} {Physical Review B}\ }\textbf {\bibinfo {volume} {62}},\ \bibinfo
  {pages} {14950} (\bibinfo {year} {2000})}\BibitemShut {NoStop}%
\bibitem [{\citenamefont {Tiberkevich}\ \emph {et~al.}(2008)\citenamefont
  {Tiberkevich}, \citenamefont {Slavin},\ and\ \citenamefont
  {Kim}}]{tiberkevich2008}%
  \BibitemOpen
  \bibfield  {author} {\bibinfo {author} {\bibfnamefont {V.~S.}\ \bibnamefont
  {Tiberkevich}}, \bibinfo {author} {\bibfnamefont {A.~N.}\ \bibnamefont
  {Slavin}}, \ and\ \bibinfo {author} {\bibfnamefont {J.}~\bibnamefont {Kim}},\
  }\href {\doibase 10.1103/PhysRevB.78.092401} {\bibfield  {journal} {\bibinfo
  {journal} {Physical Review B}\ }\textbf {\bibinfo {volume} {78}},\ \bibinfo
  {pages} {092401} (\bibinfo {year} {2008})}\BibitemShut {NoStop}%
\bibitem [{\citenamefont {Georges}\ \emph {et~al.}(2009)\citenamefont
  {Georges}, \citenamefont {Grollier}, \citenamefont {Cros}, \citenamefont
  {Fert}, \citenamefont {Fukushima}, \citenamefont {Kubota}, \citenamefont
  {Yakushijin}, \citenamefont {Yuasa},\ and\ \citenamefont
  {Ando}}]{georges2009}%
  \BibitemOpen
  \bibfield  {author} {\bibinfo {author} {\bibfnamefont {B.}~\bibnamefont
  {Georges}}, \bibinfo {author} {\bibfnamefont {J.}~\bibnamefont {Grollier}},
  \bibinfo {author} {\bibfnamefont {V.}~\bibnamefont {Cros}}, \bibinfo {author}
  {\bibfnamefont {A.}~\bibnamefont {Fert}}, \bibinfo {author} {\bibfnamefont
  {A.}~\bibnamefont {Fukushima}}, \bibinfo {author} {\bibfnamefont
  {H.}~\bibnamefont {Kubota}}, \bibinfo {author} {\bibfnamefont
  {K.}~\bibnamefont {Yakushijin}}, \bibinfo {author} {\bibfnamefont
  {S.}~\bibnamefont {Yuasa}}, \ and\ \bibinfo {author} {\bibfnamefont
  {K.}~\bibnamefont {Ando}},\ }\href {\doibase 10.1103/PhysRevB.80.060404}
  {\bibfield  {journal} {\bibinfo  {journal} {Physical Review B}\ }\textbf
  {\bibinfo {volume} {80}},\ \bibinfo {pages} {060404} (\bibinfo {year}
  {2009})}\BibitemShut {NoStop}%
\bibitem [{\citenamefont {Bianchini}\ \emph {et~al.}(2010)\citenamefont
  {Bianchini}, \citenamefont {Cornelissen}, \citenamefont {Kim}, \citenamefont
  {Devolder}, \citenamefont {Roy}, \citenamefont {Lagae},\ and\ \citenamefont
  {Chappert}}]{bianchini2010}%
  \BibitemOpen
  \bibfield  {author} {\bibinfo {author} {\bibfnamefont {L.}~\bibnamefont
  {Bianchini}}, \bibinfo {author} {\bibfnamefont {S.}~\bibnamefont
  {Cornelissen}}, \bibinfo {author} {\bibfnamefont {J.}~\bibnamefont {Kim}},
  \bibinfo {author} {\bibfnamefont {T.}~\bibnamefont {Devolder}}, \bibinfo
  {author} {\bibfnamefont {W.~v.}\ \bibnamefont {Roy}}, \bibinfo {author}
  {\bibfnamefont {L.}~\bibnamefont {Lagae}}, \ and\ \bibinfo {author}
  {\bibfnamefont {C.}~\bibnamefont {Chappert}},\ }\href {\doibase
  10.1063/1.3467043} {\bibfield  {journal} {\bibinfo  {journal} {Applied
  Physics Letters}\ }\textbf {\bibinfo {volume} {97}},\ \bibinfo {pages}
  {032502} (\bibinfo {year} {2010})}\BibitemShut {NoStop}%
\bibitem [{\citenamefont {Gusakova}\ \emph {et~al.}(2011)\citenamefont
  {Gusakova}, \citenamefont {Quinsat}, \citenamefont {Sierra}, \citenamefont
  {Ebels}, \citenamefont {Dieny}, \citenamefont {{Buda-Prejbeanu}},
  \citenamefont {Cyrille}, \citenamefont {Tiberkevich},\ and\ \citenamefont
  {Slavin}}]{gusakova2011}%
  \BibitemOpen
  \bibfield  {author} {\bibinfo {author} {\bibfnamefont {D.}~\bibnamefont
  {Gusakova}}, \bibinfo {author} {\bibfnamefont {M.}~\bibnamefont {Quinsat}},
  \bibinfo {author} {\bibfnamefont {J.~F.}\ \bibnamefont {Sierra}}, \bibinfo
  {author} {\bibfnamefont {U.}~\bibnamefont {Ebels}}, \bibinfo {author}
  {\bibfnamefont {B.}~\bibnamefont {Dieny}}, \bibinfo {author} {\bibfnamefont
  {L.~D.}\ \bibnamefont {{Buda-Prejbeanu}}}, \bibinfo {author} {\bibfnamefont
  {M.}~\bibnamefont {Cyrille}}, \bibinfo {author} {\bibfnamefont
  {V.}~\bibnamefont {Tiberkevich}}, \ and\ \bibinfo {author} {\bibfnamefont
  {A.~N.}\ \bibnamefont {Slavin}},\ }\href {\doibase 10.1063/1.3615283}
  {\bibfield  {journal} {\bibinfo  {journal} {Applied Physics Letters}\
  }\textbf {\bibinfo {volume} {99}},\ \bibinfo {pages} {052501} (\bibinfo
  {year} {2011})}\BibitemShut {NoStop}%
\bibitem [{\citenamefont {Mizushima}\ \emph {et~al.}(2009)\citenamefont
  {Mizushima}, \citenamefont {Nagasawa}, \citenamefont {Kudo}, \citenamefont
  {Saito},\ and\ \citenamefont {Sato}}]{mizushima2009}%
  \BibitemOpen
  \bibfield  {author} {\bibinfo {author} {\bibfnamefont {K.}~\bibnamefont
  {Mizushima}}, \bibinfo {author} {\bibfnamefont {T.}~\bibnamefont {Nagasawa}},
  \bibinfo {author} {\bibfnamefont {K.}~\bibnamefont {Kudo}}, \bibinfo {author}
  {\bibfnamefont {Y.}~\bibnamefont {Saito}}, \ and\ \bibinfo {author}
  {\bibfnamefont {R.}~\bibnamefont {Sato}},\ }\href {\doibase
  10.1063/1.3111435} {\bibfield  {journal} {\bibinfo  {journal} {Applied
  Physics Letters}\ }\textbf {\bibinfo {volume} {94}},\ \bibinfo {pages}
  {152501} (\bibinfo {year} {2009})}\BibitemShut {NoStop}%
\bibitem [{\citenamefont {Arai}\ \emph {et~al.}(2012)\citenamefont {Arai},
  \citenamefont {Tsukahara},\ and\ \citenamefont {Imamura}}]{arai2012}%
  \BibitemOpen
  \bibfield  {author} {\bibinfo {author} {\bibfnamefont {H.}~\bibnamefont
  {Arai}}, \bibinfo {author} {\bibfnamefont {H.}~\bibnamefont {Tsukahara}}, \
  and\ \bibinfo {author} {\bibfnamefont {H.}~\bibnamefont {Imamura}},\ }\href
  {\doibase 10.1063/1.4745777} {\bibfield  {journal} {\bibinfo  {journal}
  {Applied Physics Letters}\ }\textbf {\bibinfo {volume} {101}},\ \bibinfo
  {pages} {092405} (\bibinfo {year} {2012})}\BibitemShut {NoStop}%
\bibitem [{\citenamefont {Matsushita}\ \emph {et~al.}(2010)\citenamefont
  {Matsushita}, \citenamefont {Sato}, \citenamefont {Imamura},\ and\
  \citenamefont {Sasaki}}]{matsushita2010-a}%
  \BibitemOpen
  \bibfield  {author} {\bibinfo {author} {\bibfnamefont {K.}~\bibnamefont
  {Matsushita}}, \bibinfo {author} {\bibfnamefont {J.}~\bibnamefont {Sato}},
  \bibinfo {author} {\bibfnamefont {H.}~\bibnamefont {Imamura}}, \ and\
  \bibinfo {author} {\bibfnamefont {M.}~\bibnamefont {Sasaki}},\ }\href
  {\doibase 10.1088/1742-6596/200/4/042016} {\bibfield  {journal} {\bibinfo
  {journal} {Journal of Physics: Conference Series}\ }\textbf {\bibinfo
  {volume} {200}},\ \bibinfo {pages} {042016} (\bibinfo {year}
  {2010})}\BibitemShut {NoStop}%
\bibitem [{\citenamefont {Anh~Nguyen}\ \emph {et~al.}(2012)\citenamefont
  {Anh~Nguyen}, \citenamefont {Benatmane}, \citenamefont {Fallahi},
  \citenamefont {Fang}, \citenamefont {Mohseni}, \citenamefont {Dumas},\ and\
  \citenamefont {{\AA}kerman}}]{nguyen2012}%
  \BibitemOpen
  \bibfield  {author} {\bibinfo {author} {\bibfnamefont {T.~N.}\ \bibnamefont
  {Anh~Nguyen}}, \bibinfo {author} {\bibfnamefont {N.}~\bibnamefont
  {Benatmane}}, \bibinfo {author} {\bibfnamefont {V.}~\bibnamefont {Fallahi}},
  \bibinfo {author} {\bibfnamefont {Y.}~\bibnamefont {Fang}}, \bibinfo {author}
  {\bibfnamefont {S.~M.}\ \bibnamefont {Mohseni}}, \bibinfo {author}
  {\bibfnamefont {R.~K.}\ \bibnamefont {Dumas}}, \ and\ \bibinfo {author}
  {\bibfnamefont {J.}~\bibnamefont {{\AA}kerman}},\ }\href {\doibase
  10.1016/j.jmmm.2012.06.043} {\bibfield  {journal} {\bibinfo  {journal}
  {Journal of Magnetism and Magnetic Materials}\ }\textbf {\bibinfo {volume}
  {324}},\ \bibinfo {pages} {3929} (\bibinfo {year} {2012})}\BibitemShut
  {NoStop}%
\end{thebibliography}%

\newpage

\end{document}